# A LINEAR MIXED MODEL APPROACH FOR MEASUREMENT ERROR ADJUSTMENT: APPLICATIONS TO SEDENTARY BEHAVIOR ASSESSMENT FROM WEARABLE DEVICES

BY RUOHUI CHEN[1,a], DORI ROSENBERG[2,b] CHONGZHI DI[3,c], RONG ZABLOCKI[1,a],
SHERI J HARTMAN[1,a], ANDREA LACROIX[1,a], XIN TU[1,a], LOKI NATARAJAN[1,a], AND LIN LIU[1,a]

[1]*Herbert Wertheim School of Public Health and Human Longevity Science, University of California San Diego, CA,* [a]*9500 Gilman Dr, La Jolla, CA 92093*

[2]*Kaiser Permanente Washington Health Research Institute, Seattle, WA,* [b]*1730 Minor Ave, Seattle, WA 98101*

[3]*Division of Public Health Sciences, Fred Hutchinson Cancer Center,* [c]*1100 Fairview Ave N, Seattle, WA 98109*

In recent years, wearable devices have become more common to capture a wide range of health behaviors, especially for physical activity and sedentary behavior. These sensor-based measures are deemed to be objective and thus less prone to self-reported biases, inherent in questionnaire assessments. While this is undoubtedly a major advantage, there can still be measurement errors from the device recordings, which pose serious challenges for conducting statistical analysis and obtaining unbiased risk estimates.

There is a vast literature proposing statistical methods for adjusting for measurement errors in self-reported behaviors, such as in dietary intake. However, there is much less research on error correction for sensor-based device measures, especially sedentary behavior. In this paper, we address this gap. Exploiting the excessive multiple-day assessments typically collected when sensor devices are deployed, we propose a two-stage linear mixed effect model (LME) based approach to correct bias caused by measurement errors. We provide theoretical proof of the debiasing process using the Best Linear Unbiased Predictors (BLUP), and use both simulation and real data from a cohort study to demonstrate the performance of the proposed approach while comparing to the naïve plug-in approach that directly uses device measures without appropriately adjusting measurement errors. Our results indicate that employing our easy-to-implement BLUP correction method can greatly reduce biases in disease risk estimates and thus enhance the validity of study findings.

1

1. Introduction. Current research suggests that a sedentary lifestyle can increase potential health risks, such as all-cause mortality rate, cancer risks, and risks for metabolic diseases [1]. Many health organizations have proposed scientific advisories on sedentary behavior to encourage people to exercise and minimize their sedentary time. The World Health Organization 2020 guidelines on physical activity and sedentary behavior recommend that adults engage in 75-300 minutes of moderate to vigorous physical activities (MVPA) weekly, and reduce sedentary time [2]. The American Heart Association also issued recommendations, indicating minimizing sedentary behavior can lower cardiovascular morbidity and mortality [3]. However, compared to physical activity recommendations, guidelines for sedentary behavior have been non-specific and do not indicate how much sedentary behavior is "acceptable" nor do they quantify the amount by which sedentary behavior needs to be reduced in order to confer health benefits. In order to create specific guidelines, we have to address the measurement error inherent in current estimates of sedentary behavior. There are many sources for such measurement errors, such as inaccurate calibration of measurement instruments, inaccurate observations, recording errors. Those measurement errors can be correlated with the true value of observations, the explanatory variables, and the response variables, making an intractable obstacle to obtaining accurate assessment of sedentary behavior, and to obtaining valid estimates of associations between these health behaviors and health outcomes [4, 5, 6].

The existence of measurement errors in sedentary behavior assessment arising from different sources poses serious challenges for conducting statistical analysis and obtaining unbiased estimates, especially without validation data [7]. Many large community-based studies use surrogate tools, such as self-report for activities during the day, which are subject to recall biases and contain both random and systemic errors. While a variety of technology-based trackers for measuring physical activity and sedentary behavior have emerged over the last few decades, very few ’gold standards’ are established for evaluating the recorded measures. Failure to model the measurement errors appropriately can not only undermine the study design, but also lead to invalid conclusions, reduced statistical power, biased exposure-disease risk estimates, and misclassified risk groups [8, 9, 10, 11, 12].

Statistical methods for modeling measurement errors in dietary assessment have been intensively studied in the past decades [13, 14]. However, research to account for measurement errors in sensor-based recording activities, specifically sedentary behavior, is less studied, especially when there are multiple replicates of recordings for subjects in studies, as is typically the case for data from wearable sensors. The few published studies have focused on physical activity using regression calibration, and functional or Bayesian techniques to evaluate and correct for errors. For example, Ferrari et al (2007) [15], Nusser et al (2012) [16], and Beyler et al (2013) [17] proposed a multi-level equation-based modeling process to evaluate different types of physical activity measures and estimate the validity coefficients and attenuation factors, while accounting for measurement errors. Lim et al (2015) [18] proposed using regression calibration method to account for measurement errors in a self-reported physical activity survey in New York city. Agogo et al (2015) [19] and Jadhav et al (2022) [20] proposed a Bayesian-based method and a function-based method to model measurement error for physical activity data. Morrell et al. (2003) [21] proposed a method using the LME model on repeated measurements to obtain a predicted value to account for errors in the measurements. Our approach also exploits repeated measures but our focus is on analytic derivations to prove the unbiasedness of our proposed estimates. To our knowledge, no studies to date have evaluated measurement error for sedentary behavior derived from sensors, with a focus on providing statistical calculations, as well as, real data applications, of the process for correcting bias caused by measurement errors.

While theoretical approaches for measurement error correction could in principle be transported from one setting (e.g., diet) to another (e.g., sensor-based sedentary behavior), it is necessary to conduct a careful evaluation of the unique measurement properties of a given device and health behavior, in order to develop rigorous and domain-specific measurement correction tools. In this article, we conduct such an analysis of sedentary behavior derived from two commonly used wearable sensors (ActiGraph and activPAL). Leveraging the availability of multiple replicates measured from both devices for each subject, we propose to use structure models consisting of Linear Mixed Effect Models and Generalized Linear Models to obtain unbiased estimates of the relationship between exposures subject to measurement errors and outcome of interest, after appropriately accounting for the errors in devices' measurement. Section 2 introduces the sedentary behavior study that provides the motivation for the current work. Section 3 introduces the structure models to appropriately account for measurement errors and the proof of measurement error correction process. Section 4 implements Monte Carlo simulation to evaluate the proposed method. In Section 5 we apply our proposed method to the sedentary behavior study. Section 6 discusses the contributions and limitations of the proposed method.

## 2. Motivating Data Example.

2.1. *Study Cohort.* Our proposed method and data application were motivated by the Adult Changes in Thought(ACT) study, which is an ongoing longitudinal cohort study of community-dwelling older adults that were greater than 65 years old and without evidence of Alzheimer's Disease or dementia in Washington State. The ACT study was initiated in 1994 to investigate risk factors for the development of dementia and has since provided a unique set of opportunities to additionally study a wide range of non-cognitive factors of healthy aging. Pertinent to the current study, the ACT activity monitor sub-study (ACT-AM) was initiated in 2016, adding a device-based activity component to capture the spectrum of sedentary and physical activity patterns [22]. Participants were instructed to wear a hip-worn triaxial ActiGraph GT3X+ (ActiGraph LLC, Pensacola, FL, USA) which captured 30Hz movement accelerations in three spatial axes, and a thigh-worn activPAL micro3 (PAL Technologies, Glasgow, Scotland, UK), which captured postures (e.g., sitting, standing, moving) [23]. Participants were instructed to wear both devices at the same time for 7 days. Participants also recorded self-reported sleep logs, in which participants recorded the time of waking up and going to bed, throughout their device wear. Ethics approval was obtained from the Kaiser Permanente Washington institutional review board. All participants provided written informed consent. There were total 980 subjects included in the data analysis; Table 1 provides demographic summary statistics of the study sample.

2.2. *Sedentary Behavior Measures.* Our data comprised concurrent wear of two devices, namely thigh-worn activPAL and hip-worn ActiGraph. The thigh-worn activPAL has been widely used and is considered a gold standard for measuring postures, specifically sitting. Minutes spent in a sitting or lying posture during waking hours were summed over the day to provide a daily total sedentary time estimate for the activPAL [24]. For the ActiGraph monitor, sedentary behavior is commonly estimated using cut-points; we used the standard cut-point of <100 counts per minute for capturing participant's daily total sedentary time [25]. More than 95 % of subjects have at least 5 days of wearing both devices in the study, indicating the percentage of missing days of wearing devices for participants is very low.

We compared estimates of total daily sedentary time over the wearing period for each subject from these two devices via boxplots and summary statisics. From Fig 1 and Table 2, we can see that the daily average sitting time from activPAL is slightly lower than the same recording from ActiGraph with higher variance. The longer sitting time measured by the ActiGraph could be caused by measurement error inherent when using cut-points to delineate (in)activity. Meanwhile, the Pearson correlation between measures from activPAL and

Table 1

*Descriptive Characteristics Of The Study Cohort*

| Characteristics | Subjects Included (n=980) |
|---|---|
| Age in years, mean(SD) | 77.0(6.6) |
| Male,n(%) | 428(44.7) |
| Race/Ethnicity, n(%) Hispanic or non-White | 100(10.2) |
| non-Hispanic White | 876(89.8) |
| Education, n(%) Less than high school | 15(1.5) |
| Completed high school | 79(8.1) |
| Some College | 157(16.0) |
| Completed College | 729(74.4) |
| BMI, n(%) Underweight (≤18.5) | 8(0.8) |
| Normal (18.5 - 24) | 357(37.2) |
| Overweight (25 - 29) | 378(39.4) |
| Obese (≥30) | 216(22.5) |

Fig 1. *Boxplot of Daily Average Total Sedentary Time (mins) from activPAL and ActiGraph*

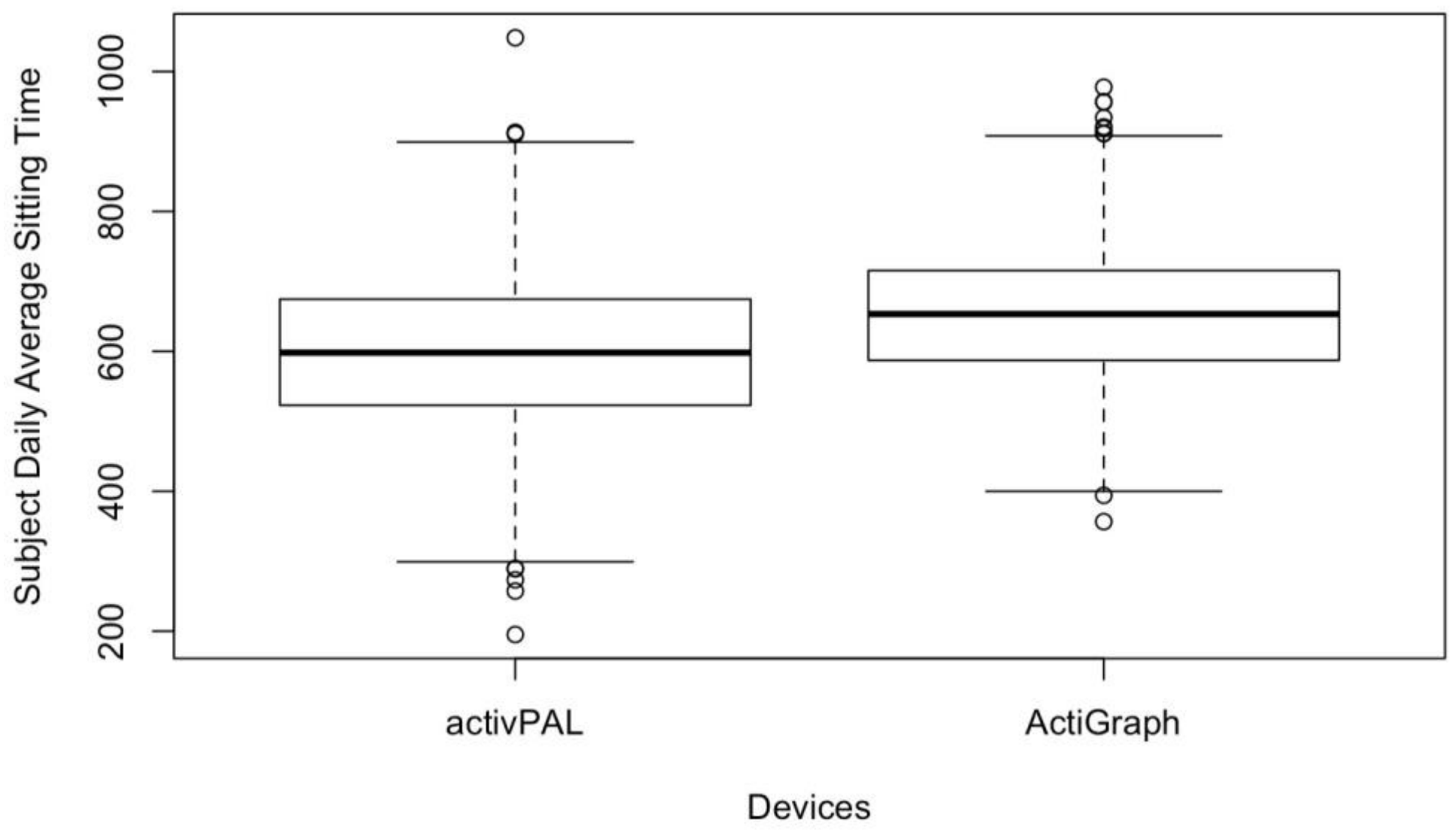

TABLE 2
*Daily Average Sedentary Time (mins) from activPAL and ActiGraph*

| Device (mins) | Min | Median | Mean(SD) | Max |
|---|---|---|---|---|
| activPAL | 195.2 | 598.6 | 598.6(120.5) | 1048.3 |
| ActiGraph | 356.6 | 655.9 | 654.9(97.6) | 977.6 |

measures from ActiGraph is around 0.64 with a p-value less than 0.001, indicating that there is a significant positive relationship between measures from the two devices.

Fig 2 is the scatter plot between measures from activPAL and measures from ActiGraph, the blue line is added as the reference for 'y=x', representing the perfect agreement between the two device estimates. Each dot represents a subject. We can see that the majority of the dots are above the blue reference line, indicating that most of the measures from activPAL are smaller than the measures from ActiGraph. Besides the scatter plot, we also used the BlandAltman plot to investigate the concordance between measures from the two devices [26]. In Fig.3, the blue dotted line indicates the mean difference measures between the two devices, and the light blue dashed line indicates the 95% confidence interval for the difference. From the Bland-Altman plot, we can see that the mean measures from activPAL are slightly smaller than the mean measures from ActiGraph, while many of the points are spread over the 95% confidence interval bands, indicating substantial variability in agreement between the two devices.

In summary, the plots and summary statistics clearly indicate that sedentary behavior estimates from the two devices are not identical. Importantly, even though the activPAL is considered to be more accurate for capturing sedentary behavior compared to the ActiGraph, measures from both devices likely contain measurement errors, which are due to inaccurate device calibration, inappropriate device wearing styles, recording errors. Naïvely using the measures without appropriately accounting for the measurement errors can give us biased estimates and ultimately lead to invalid conclusions. We investigate these issues in the next sections, and propose using structure models to account for the measurement errors, with the ultimate goal of obtaining unbiased and consistent estimates.

FIG 2. *Total Sedentary Time estimates between activPAL and ActiGraph*

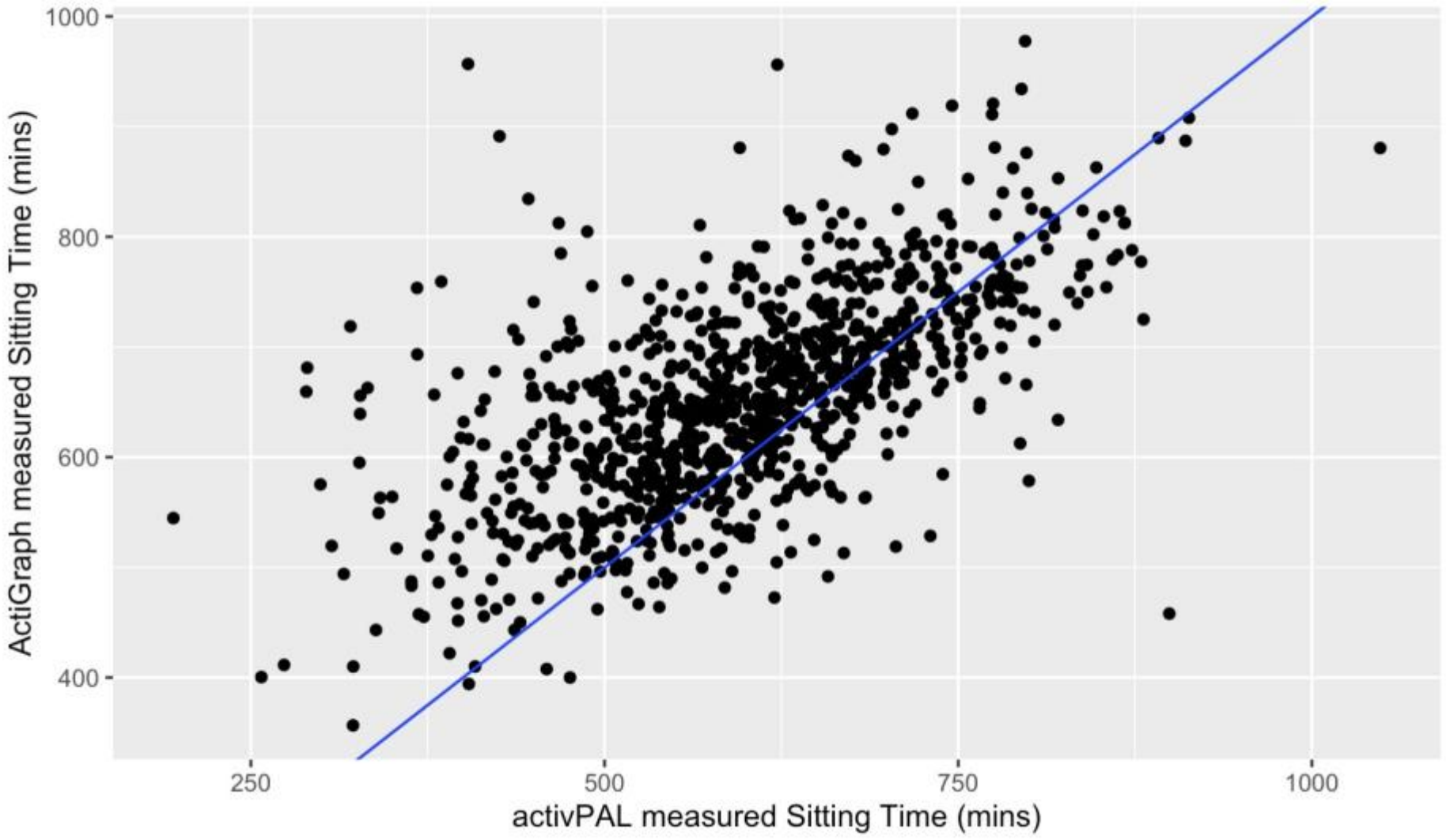

3. Structure Models.

3.1. *Model Setup.* We first establish terminology and the models. Let $n$ denote the number of subjects, and assume every subject has the same number of replicates $J$. Consider our measurement error and outcome models as below:

$$W_{ij} = \sum_{s=1}^{m} \gamma_{0s} A_{is} + \gamma_1 X_i + U_{ij} \tag{1}$$

$$Y_i = \beta_0 + \beta_x X_i + \sum_{c=1}^{p} \beta_c C_{ic} + \epsilon_i \tag{2}$$

FIG 3. *Total Sedentary Time Difference between activPAL and ActiGraph*

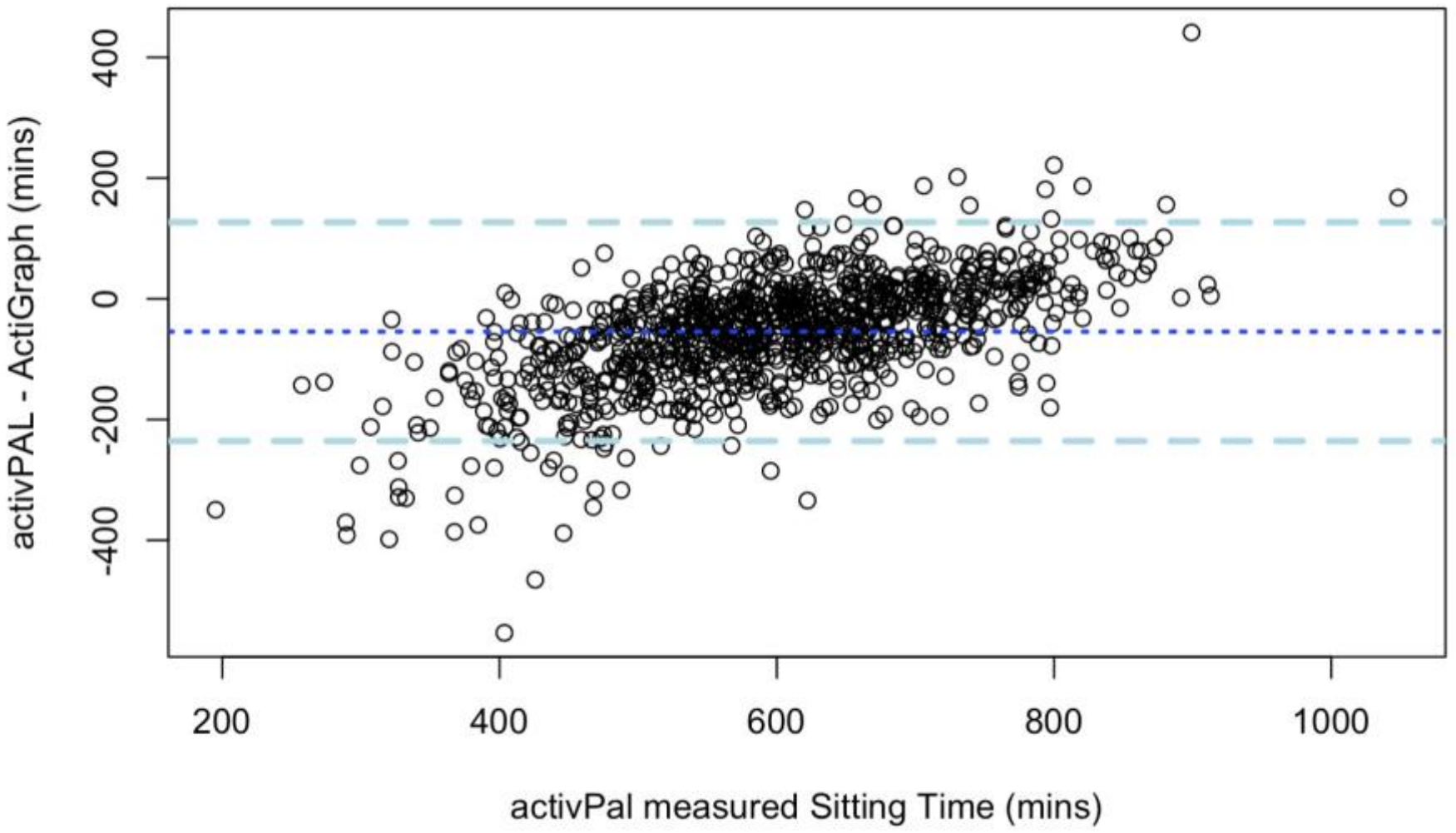

where $W_{ij}$ denotes observed sedentary behavior measures for subject $i$ from a device, with $j$ represents repeated measures. $A_{is}$ denotes a covariate that is without measurement error and informative for the measures of $W_{ij}$ for subject $i$ and there are a total of $m-1$ such covariates, with $A_{i1}$ equals 1 for all subjects to include an intercept in the model. $C_{ic}$ denotes a covariate without measurement error that is correlated with an outcome measure $Y_i$, $p$ denotes the number of such covariates for subject $i$. Meanwhile $X_i$ represents the centered true measures without measurement error with $X_i \sim N(0, \sigma_x^2)$, $\epsilon_i \sim N(0, \sigma_\epsilon^2)$. $\gamma_1$ represents the attenuation bias when we use surrogate $W$ instead of the true $X$. $\beta_x$ is the coefficient of interest and represents the association between the true exposure, X, and outcome Y.

We introduce some matrix notations next to simplify the analytic derivations to follow. Let $X = (X_1, X_2, ..., X_n)^T_{1\times n}$, $\boldsymbol{Y} = (Y_1, Y_2, ..., Y_n)^T_{1\times n}$, and $\epsilon = (\epsilon_1, \epsilon_2, ..., \epsilon_n)^T_{1\times n}$.

Let $U_i$ denote a vector of the errors $U_{ij}$ for subject $i$ and $U$ denote a vector of $U_i$ and, then $U_i = (U_{i1}, U_{i2}, ..., U_{iJ})^T_{1\times J} \sim N(0, \boldsymbol{\Sigma_u})$ and $U = (\boldsymbol{U_1}, \boldsymbol{U_2}, ..., \boldsymbol{U_n})^T_{1\times nJ} \sim N(0, \boldsymbol{\Sigma})$, where

$$\boldsymbol{\Sigma} = \begin{pmatrix} \boldsymbol{\Sigma}_u & & & \\ & \boldsymbol{\Sigma}_u & & \\ & & \ddots & \\ & & & \boldsymbol{\Sigma}_u \end{pmatrix}_{nJ\times nJ}$$

with

$$\boldsymbol{\Sigma_u} = \begin{bmatrix} \sigma_u^2 & \rho\sigma_u^2 \cdots & \\ & \sigma_u^2 \cdots & \\ & & \sigma_u^2 \cdots \end{bmatrix}_{J\times J},$$

where $\rho$ denotes the error correlation between replicates within each subject for the device, such as the error correlation between $U_{ij}$ and $U_{ik}$ ($j$ and $k$ represent different measures for the same subject $i$). Note we are assuming error correlations are exchangeable which is reasonable in our case, since the measures are collected in close proximity, e.g., daily over 7 days.

Let $C$ denote a matrix of $C_{ic}$ so that $C = \begin{pmatrix} \boldsymbol{C_1} \\ \boldsymbol{C_2} \\ \vdots \\ \boldsymbol{C_n} \end{pmatrix}_{n\times p}$, where $C_i = (C_{i1}, ..., C_{ip})_{1\times p}$ indicates a vector of covariates without measurement errors for subject $i$; and let $\beta_c$ denote a vector of parameters $\beta_c$ that is the coefficient for each covariate without measurement error $\boldsymbol{\beta_c} = (\beta_1, \beta_2, ..., \beta_p)^T_{1\times p}$.

Let $A$ denote a matrix that

$$\begin{pmatrix} \boldsymbol{\tilde{A}_{11}} \\ \vdots \\ \boldsymbol{\tilde{A}_{1J}} \end{pmatrix}$$

$$A = \begin{pmatrix} \vdots \\ \tilde{A}_{21} \\ \vdots \\ \tilde{A}_{2J} \\ \vdots \\ \tilde{A}_{n1} \\ \vdots \\ \tilde{A}_{nJ} \end{pmatrix}_{nJ \times m}$$

where $\tilde{A}_{ij} = (A_{i1}, A_{i2}, ..., A_{im})$ with the first column being constant 1 to incorporate intercept. Let $\gamma_0$ denote a vector

$$\gamma_0 = (\gamma_{01}, \gamma_{02}, ..., \gamma_{0m})^T_{1\times m},$$

and $W$ denote a vector of the replicates $W_{ij}$ for subject $i$

$$W = (W_{11}, .., W_{1J}, W_{21}, .., W_{2J}, .., W_{n1}, .., W_{nJ})^T_{1\times nJ}$$

We also define $Z_i = \mathbf{1}_{J\times 1}$, a vector of length $J$ of 1s, and $Z$ is a matrix of $Z_i$

$$Z = \begin{pmatrix} Z_1 & & & \\ & Z_2 & & \\ & & \ddots & \\ & & & Z_n \end{pmatrix}_{nJ\times n}$$

Then we can re-write the measurement error model (1) and outcome model (2) as

$$W = A\gamma_0 + Z\gamma_1 X + U \tag{3}$$

$$Y = \beta_0 \mathbf{1}_{n\times 1} + \beta_x X + C\beta_c + \epsilon \tag{4}$$

Therefore, $\gamma_1 \boldsymbol{X} \sim N(0, \gamma_1^2 \sigma_x^2 \boldsymbol{I_n})$, where $I_n$ is a $n \times n$ identity matrix. For notational brevity, let $G = Var(\gamma_1 \boldsymbol{X}) = \gamma_1^2 \sigma_x^2 \boldsymbol{I_n}$. Then, using standard linear mixed effects (LME) model theory [27], we can estimate $\gamma_0$ in the measurement error model by $\hat{\gamma}_0 =$ $(A^TV^{-1}A)^{-1}A^TV^{-1}W$, where $V$ represents the variance for $W$ given $A$ with $V = ZGZ^T + \mathbf{\Sigma}$. We can then estimate the Best Linear Unbiased Predictors (BLUP) from the measurement error model (3), $\widehat{\gamma_1 \boldsymbol{X}} = GZ^TV^{-1}(W - A\hat{\gamma}_0)$.

3.2. *Measurement Error Correction and Unbiasedness.* It is well-known that naïvely plugging in a replicate value, e.g., $W_{ij}$ or the average across replicates as a substitute for the true measures will usually give biased estimates of $\beta_x$ [9]. We will demonstrate this point for our application in a later section, here we prove that using the BLUP (from the LME model) in the outcome model gives us unbiased estimators for $\beta_x$ when the outcome is continuous. In a subsequent section, we also evaluate the performance of the proposed method for both continuous and binary outcomes for different sample sizes through simulations.

Let $H = I_{nJ} - A(A^TV^{-1}A)^{-1}A^TV^{-1}$, where $I_{nJ}$ is a $nJ \times nJ$ identity matrix. Note that $H$ is idempotent since $H^2 = H$, and that $V^{-1}H = H^TV^{-1}$. We also have $HA\gamma_0 = \mathbf{0}$, where $\mathbf{0}$ is a $nJ \times 1$ matrix (vector), and therefore $HW = H(Z\gamma_1X + U)$.
Therefore, the estimated $\gamma_1X = GZ^TV^{-1}HW = GZ^TV^{-1}H(Z\gamma_1X + U)$.

The joint distribution of $Y$ and $\gamma_1X$ is multivariate normal, assuming independence between $X$ and $C$, we can write:

$$\begin{bmatrix} \boldsymbol{Y} \\ \widehat{\gamma_1 \boldsymbol{X}} \end{bmatrix} = \begin{bmatrix} \beta_0 \mathbf{1}_{n\times 1} & \beta_x I_n & C I_n & 0 & 0 \\ 0 & GZ^TV^{-1}HZ\gamma_1 & 0 & 0 & GZ^TV^{-1}H \end{bmatrix} \begin{bmatrix} 1 \\ \boldsymbol{X} \\ \boldsymbol{\beta}_c \\ \boldsymbol{\epsilon} \\ U \end{bmatrix}$$

.

$$Var(\widehat{\gamma_1 \boldsymbol{X}}) = Var((GZ^TV^{-1}H)(Z\gamma_1X + U))$$
$$= GZ^TV^{-1}H[Var(Z\gamma_1X + U)]H^T(V^{-1})^TZG^T$$
$$= GZ^TV^{-1}HVH^TV^{-1}ZG^T$$
$$= GZ^TH^TV^{-1}ZG^T$$

$$Cov(Y, \widehat{\gamma_1 X}) = Cov(\beta_0\mathbf{1}_{n\times 1} + \beta_xX + C\beta_c + \epsilon, GZ^TV^{-1}H(Z\gamma_1X + U))$$
$$= Cov(\beta_xX, GZ^TV^{-1}HZ\gamma_1X)$$
$$= \beta_xCov(X, \gamma_1X)(GZ^TV^{-1}HZ)^T$$
$$= \beta_x\gamma_1Var(X)Z^TH^T(V^{-1})^TZG^T$$
$$= \frac{\beta_x}{\gamma_1}Var(\gamma_1\boldsymbol{X})Z^TH^TV^{-1}ZG^T$$
$$= \frac{\beta_x}{\gamma_1}GZ^TH^TV^{-1}ZG^T$$

Then by the properties of multivariate normal distribution, we have:

$$E(\boldsymbol{Y}|\widehat{\gamma_1\boldsymbol{X}},\boldsymbol{C}) = E(\boldsymbol{Y}|\boldsymbol{C}) + Cov(\boldsymbol{Y},\widehat{\gamma_1\boldsymbol{X}})[Var(\widehat{\gamma_1\boldsymbol{X}})]^{-1}(\widehat{\gamma_1\boldsymbol{X}} - E(\widehat{\gamma_1\boldsymbol{X}}))$$

$$= \beta_0\mathbf{1}_{n\times 1} + C\beta_c + \frac{\beta_x}{\gamma_1} GZ^THTV^{-1}ZG^T\left(GZ^THTV^{-1}ZG^T\right)^{-1}(\widehat{\gamma_1\boldsymbol{X}} - 0)$$

$$= \beta_0\mathbf{1}_{n\times 1} + C\beta_c + \frac{\beta_x}{\gamma_1}\widehat{\gamma_1 X}$$

$$= (\mathbf{1}_{n\times 1}, \widehat{\boldsymbol{X}}, \boldsymbol{C})\begin{pmatrix}\beta_0\\ \beta_x\\ \boldsymbol{\beta_c}\end{pmatrix}$$

Let $D$ denote the design matrix, which is $D = (\mathbf{1}_{n\times 1}, X_C, C)$. By the ordinary least square estimation, we have:

$$\widehat{\boldsymbol{\beta}} = \begin{pmatrix}\beta_0\\ \beta_x\\ \boldsymbol{\beta_c}\end{pmatrix} = (D^TD)^{-1}D^TY$$

Then we show that $(D^TD)^{-1}D^TY$ is an unbiased estimator for $\beta_0$, $\beta_x$ and $\beta_c$:

$$E(\widehat{\boldsymbol{\beta}}) = (D^TD)^{-1}\boldsymbol{D^T}E(\boldsymbol{Y}|\widehat{\gamma_1\boldsymbol{X}},\boldsymbol{C})$$

$$= (D^TD)^{-1}D^TD\beta_b$$

$$= \beta_b$$

Therefore using the BLUP of $\gamma_1 X$, instead of naïvely observed measures which contain errors, gives us unbiased estimates for parameters in the outcome model. However, the above proof holds if and only if $C$ is independent of $X$, namely that the error-free covariates are independent of the true measures for each subject. We explored situations when $C$ and $X$ are correlated for both continuous and binary outcomes $Y$ through simulations in the next section.

4. Simulation.

4.1. *Simulation Model Setup.* Recapitulating the earlier notation, let $Y_i$ represent a clinical outcome of interest for subject $i$, $C_i$ denote a covariate measured without error, and $W_{ij}$ represent observed sedentary behavior measures from a device, where $j$ indexes replicate measures (e.g., $j = 1,2,\cdots 7$ if the device is worn daily for 1 week). $X_i$ represents the true centered average sedentary time for subject $i$ over the measurement period (e.g., 1 week). For demonstration purposes, we consider a regression model of the clinical outcome $Y_i$ on covariate $X_i$ through a link function $f(y)$. We specify the following measurement error model for the covariate measured with error,

$$W_{ij} = \gamma_0 + \gamma_1 X_i + U_{ij}; \quad (5)$$

and we specify an outcome model for the centered true average measure given the covariate $C_i$:

$$f(Y_i) = \beta_0 + \beta_x X_i + \beta_c C_i + \epsilon_i. \tag{6}$$

The equations (5) and (6) are simplified cases of (1) and (2) with no covariates $A_i$ and a single covariate $C_i$ when we consider a linear regression of a continuous outcome $Y_i$. The goal of the analysis is to estimate $\beta_x$, the effect of covariate $X$ with measurement error.

Let $W_i = (W_{i1}, W_{i2}, \cdots, W_{iJ})^T$ and $U_i = (U_{i1}, U_{i2}, \cdots, U_{iJ})^T$ be vectors representing the $J$ replicates for subject $i$, we assume

$$U_i \sim N(0, \boldsymbol{\Sigma_u}), \quad \epsilon_i \sim N(0, \sigma_\epsilon^2), \; W_i \mid X_i \sim N(\gamma_0 + \gamma_1 X_i, \boldsymbol{\Omega}),$$

$$\boldsymbol{\Omega} = \gamma_1^2 \sigma_x^2 \boldsymbol{I}_{J\times J} + \boldsymbol{\Sigma_u}, \quad \boldsymbol{\Sigma_u} = \begin{pmatrix} \sigma_u^2 & \rho\sigma_u^2 & \ldots \\ & \sigma_u^2 & \ldots \\ & & \sigma_u^2 \ldots \end{pmatrix}_{J\times J.}$$

$X$ and $C$ will be generated as follows:

$$\begin{pmatrix} X \\ C \end{pmatrix} = N\left(\begin{pmatrix} \mu_x \\ \mu_c \end{pmatrix}, \Sigma_{xc}\right), \quad \Sigma_{xc} = \begin{pmatrix} \sigma_x^2, & \rho_{xc}\sigma_x\sigma_c \\ \rho_{xc}\sigma_x\sigma_c, & \sigma_c^2 \end{pmatrix}.$$

In the above model set-up, $\rho$ denotes the error correlations between replicates within each subject for the device, $\rho_{xc}$ represents the correlation between the latent variable $X$ and variable without measurement error $C$, $N()$ indicates the normal distribution and $I_{J\times J}$ is a $J \times J$ identity matrix. We also note that for simplicity we assume that the number of replicates $J$ is the same across subjects; our methods will be easily generalized to the unbalanced case.

We conducted a series of simulations to examine the performance of the proposed method for accounting for measurement error in continuous and binary outcome models, and compared the proposed method to the naïve method that uses the error-prone measures without adjusting measurement errors. All simulations were performed with a Monte Carlo sample of 1000. We examined the performance of these methods in estimating $\beta_0$, $\beta_x$ and $\beta_c$. The performance metrics include mean estimates of the coefficient, estimated asymptotic and empirical standard errors, relative bias and coverage probability of 95% confidence interval . To mimic the dataset in the ACT-AM study, all parameters in the simulation are set based on prior analysis. Using data structures as in (5) and (6) , we showed the simulation results for sample sizes $n = 50, 100, 500$ for continuous outcome, and sample sizes $n = 100, 200, 500$ for binary outcomes; each subject had J=7 replicates of measures.

We also conducted simulations with missing completely at random for replicates of each subject, and the results are similar to using complete data.

4.2. *Continuous Clinical Outcome.* We considered a linear regression for a continuous outcome $Y_i$ with $f(y) = y$, then

$$Y_i|(X_i, C_i) \sim N(\beta_0 + \beta_x X_i + \beta_c C_i, \sigma_\epsilon^2).$$

Parameters and generated measures were as follows:

$\beta_0 = 10$, $\beta_x = 2.95$, $\beta_c = 3$, $\gamma_0 = 1$, $\gamma_1 =$ $1 (W_1)$ *or* $2 (W_2)$, $\mu_c = 1$, $\sigma_c = 1$, $\mu_x = 0$, $\sigma_x = 2$,

$\sigma_\epsilon = 1$, $\sigma_u = 1$

$\rho = 0.1$ or $0.3$, $\rho_{xc} = 0$ or $0.5$

For the correlation between the latent variable $X$ and variable without measurement error $C$, we explored the cases when $\rho_{xc} = 0$ and $\rho_{xc} = 0.5$, meanwhile varied the value of attenuation bias $\gamma_1$, where $W_1$ represents an unbiased case with measurements with $\gamma_1 = 1$, and $W_2$ represents measurements with $\gamma_1 = 2$.

4.3. *Binary Clinical Outcome.* We considered a logistic regression for a binary outcome

$$Y_i|(X_i, C_i) \sim Bernoulli(Pr), \quad Pr = \frac{1}{1 + exp(-(\beta_0 + \beta_x X_i + \beta_c C_i))}.$$

Parameters in the outcome model ($\beta$) were set up as follows, other parameters were set up the same as in continuous case:

$$\beta_0 = 0.1, \qquad \beta_x = 0.1, \qquad \beta_c = 0.1$$

FIG 4. *Relative bias of estimated $\beta_x$ using different methods for continuous outcome and binary outcome*

Fig 4.a continuous outcome with rho = 0.1 and rho_{xc} = 0

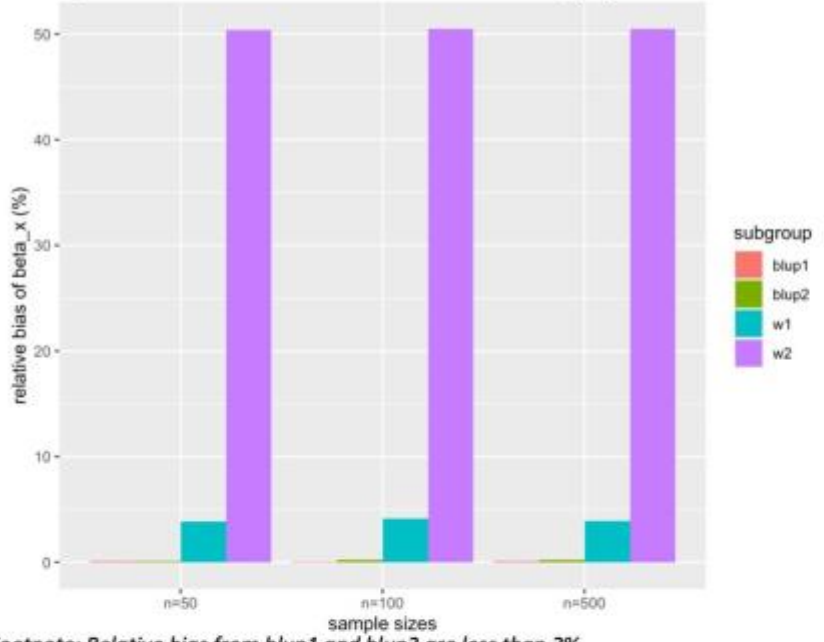

*Footnote: Relative bias from blup1 and blup2 are less than 2%*

Fig 4.b binary outcome with rho = 0.1 and rho_{xc} = 0

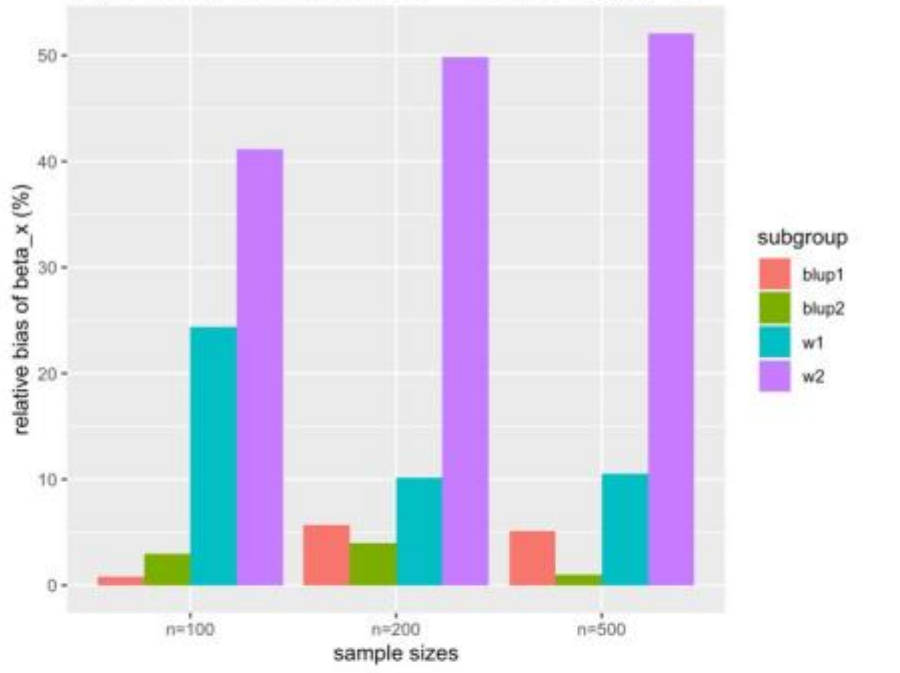

Fig 4.c continuous outcome with rho = 0.1 and rho_{xc} = 0.5

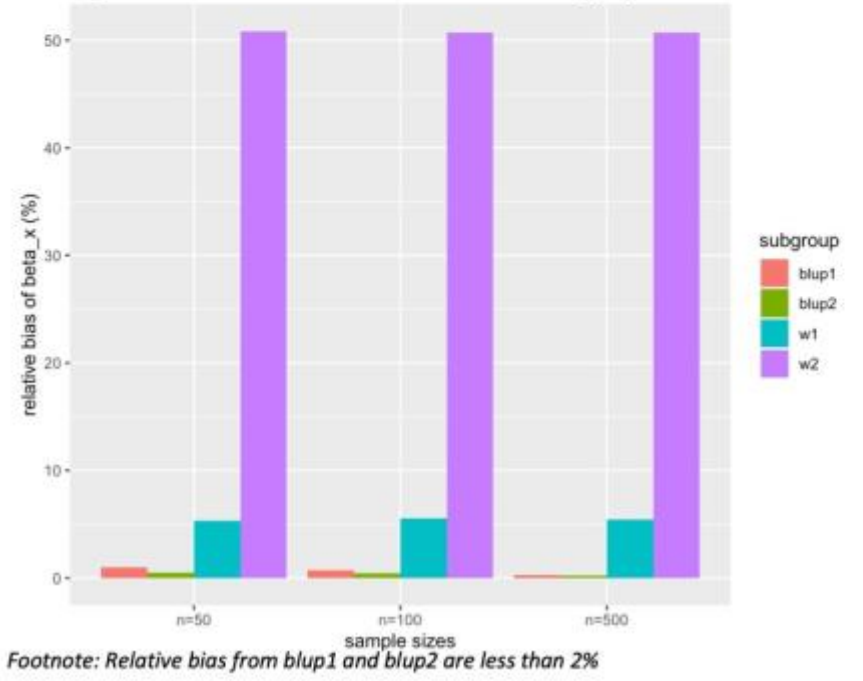

*Footnote: Relative bias from blup1 and blup2 are less than 2%*

Fig 4.d binary outcome with rho = 0.1 and rho_{xc} = 0.5

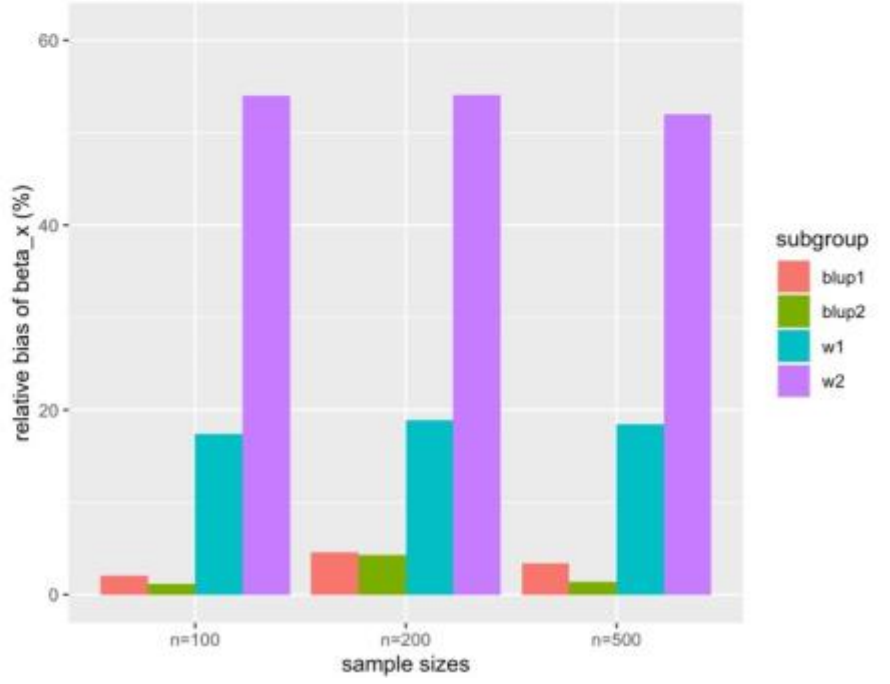

Fig 4.e continuous outcome with rho = 0.3 and rho_{xc} = 0

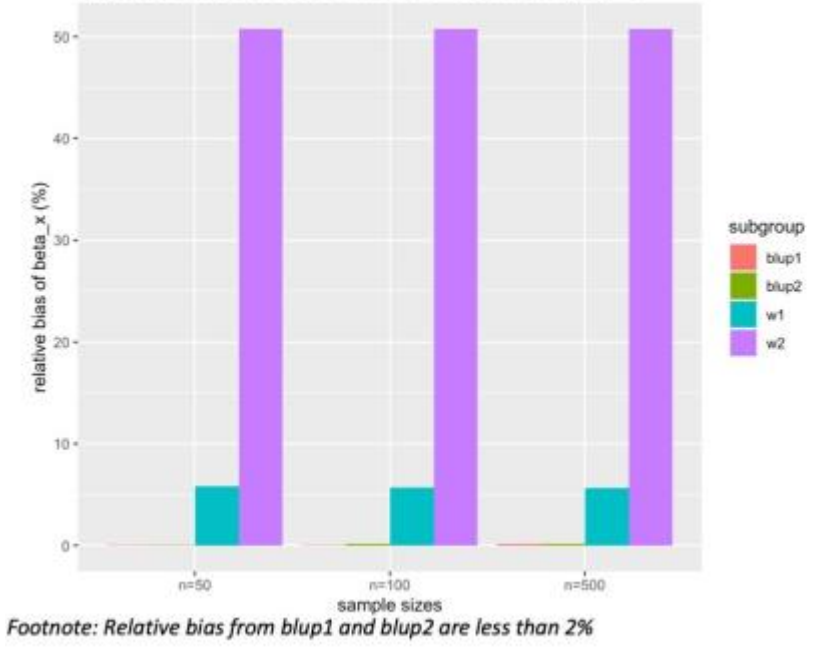

*Footnote: Relative bias from blup1 and blup2 are less than 2%*

Fig 4.f binary outcome with rho = 0.3 and rho_{xc} = 0

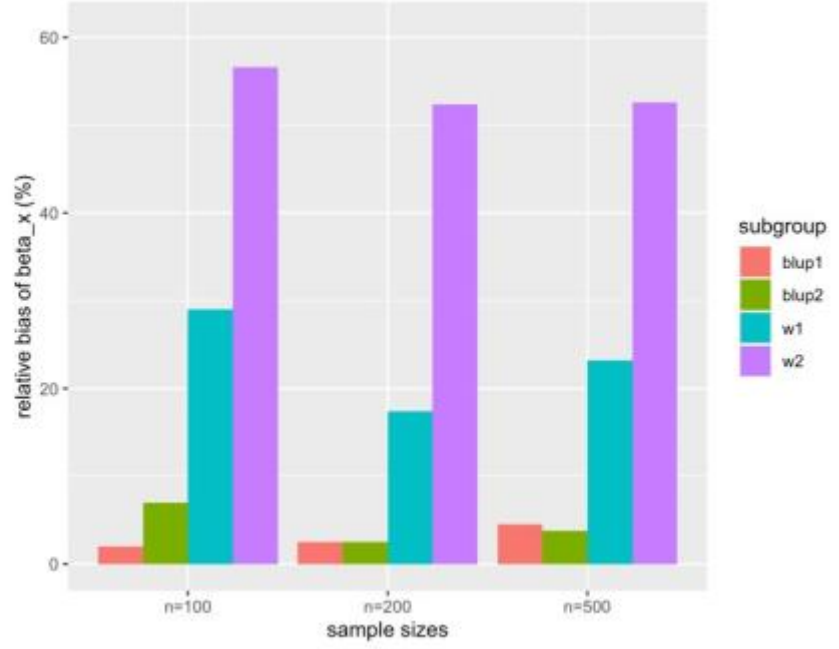

Fig 4.g continuous outcome with rho = 0.3 and rho_{xc} = 0.5

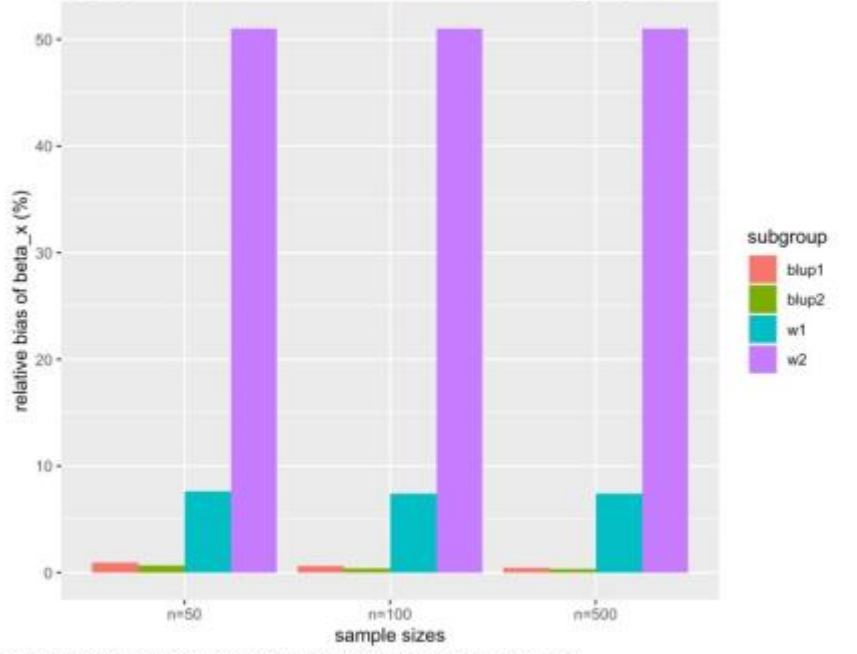

*Footnote: Relative bias from blup1 and blup2 are less than 2%*

Fig 4.h binary outcome with rho = 0.3 and rho_{xc} = 0.5

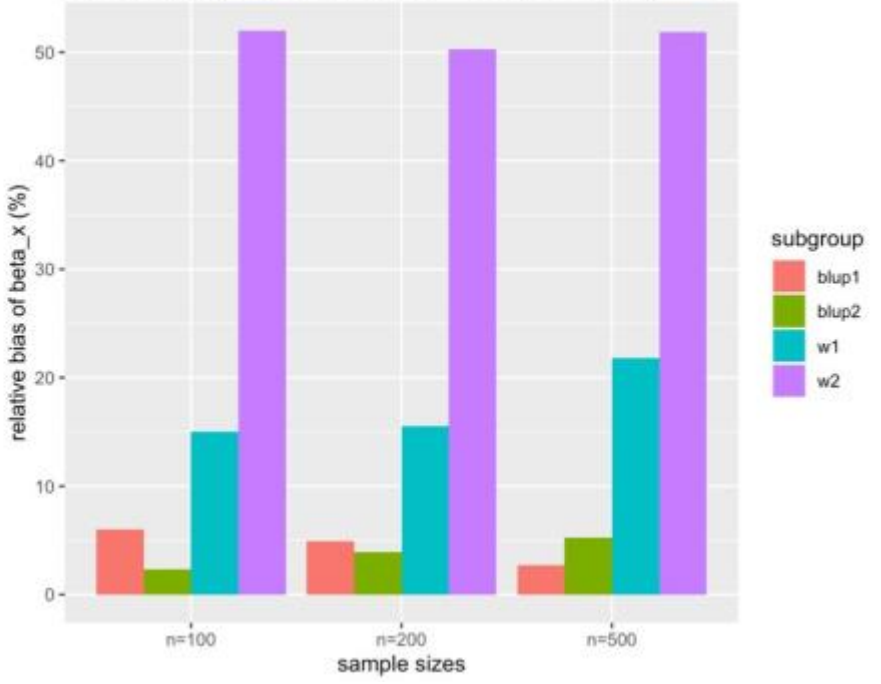

4.4. *Simulation Results.* We compared the performance of the method naïvely using the measures containing errors (denoted as $W_1$ and $W_2$ assuming the attenuated bias $\gamma_1$ =1 and 2) and the proposed approach using BLUP under 1000 Monte Carlo samples. The $BLUP_1$ and $BLUP_2$ represent using the Best Linear Unbiased Prediction (BLUP) from $W_1$ and $W_2$, respectively. Shown in the appendix Table 4 and Table 5 are the simulation results for the estimated $\beta_0$, $\beta_x$, and $\beta_c$ and standard errors of these estimates for the continuous and binary outcome models, respectively. For both continuous and binary outcomes, the standard errors of the estimates (i.e., the asymptotic standard errors) from the proposed approach using BLUP were very similar to their empirical standard errors, and as expected, these standard errors decreased as the sample size increased. In Fig 4. (Appendix Table 6 and Table 7), we summarized the relative bias of estimated $\beta_x$ under different sample sizes while comparing using $BLUP_1$ and $BLUP_2$ to naively using W1 and W2 for both continuous and binary outcomes with different magnitudes of $\rho$ and $\rho_{xc}$. As we can see from Fig 4., the proposed BLUP method has a much smaller relative bias than naively using measures containing errors for both continuous and binary outcomes under different correlations.

When the latent truth $X$ is independent of the covariate $C$ ($\rho_{xc} = 0$), the estimates of $\beta_0$, $\beta_x$ and $\beta_c$ from the proposed BLUP approach were close to the truth for both the continuous and binary outcome cases, even for relatively small sample size. In contrast, naïvely using measures containing measurement errors gave us biased estimates for $\beta_0$ and $\beta_x$, and the bias increases as the correlation for replicates within subject $\rho$ and the magnitude of attenuation bias $\gamma_1$ increases. The relative bias was given in Fig 4. For example, in the continuous outcome case, when $\rho$ = $0.1$ and $\rho_{xc}$ = $0$, the relative bias of estimated $\beta_x$ is around $0.1\%$ for rather small sample size $n = 50$ using $BLUP_1$ and $BLUP_2$ , but is around $3.8\%$ if naïvely using $W_1$ and even worse around $50\%$ if naïvely using $W_2$, which has greater attenuation bias than $W_1$ (Fig 4.a). In the binary outcome case, a similar trend follows but with a slightly increased relative bias for all methods in comparison; taking the $\rho = 0.1$ and $\rho_{xc} = 0$ for binary outcome case for example, the relative bias of estimated $\beta_x$ is around $0.8\%$ and $3.0\%$ for n=100 when using $BLUP_1$ and $BLUP_2$, in comparison to a much higher $24.4\%$ and $41.2\%$ when naïvely using $W_1$ and $W_2$ (Fig 4.b). Of note, in our simulations, we set $\gamma_1$ = 2 for $W_2$, which resulted in attenuation of $\beta_x$ when naïve plugging in $W_2$. A different choice of $\gamma_1$ (e.g., $\gamma < 1$) would result in inflation of $\beta_x$, namely erroneous exaggerated effects of the exposure-disease associations.

When $X$ is correlated with $C$ ($\rho_{xc} = 0.5$), the proposed structure models using BLUP were still able to have much less biased estimates of $\beta_0$ and $\beta_x$ than the naïve plug-in approach. For example, in continuous outcome case, when $\rho$ = $0.1$ and $\rho_{xc}$ = $0.5$, the relative bias of estimated $\beta_x$ is around $1.0\%$ and $0.5\%$ when using $BLUP_1$ and $BLUP_2$, which are much lower than $5.3\%$ and $50.9\%$ the relative bias of estimated $\beta_x$ for n=50 when naïvely using $W_1$ and $W_2$ (Fig 4.c). A similar trend was also observed in binary outcome cases with slightly increased relative bias.

Besides that, we have also examined the $95\%$ CI coverage probability for using $BLUP_1$, $BLUP_2$, $W_1$, and $W_2$, and found that for continuous outcome the coverage of the proposed method using BLUP is much better than using the naïve plug-in approach (Appendix Table 6 and Table 7). On the other hand, for binary outcome, due to the wide confidence interval of the estimates, coverage probability is similar among all approaches, but as attenuation bias $\gamma_1$ contained in the measurements increases, the coverage drops for the naïve plug-in approach, while the BLUP still maintained good coverage adjusting the measurement errors.

5. Data Analysis. In sedentary behavior studies, research generally focuses on whether an individual's activity level or sitting time has an impact on their health. We implemented our proposed method on data from the ACT Study, to investigate the impacts of increasing subjects' daily total sitting time on BMI and obesity status (i.e., BMI is equal to or greater than 30 $kg/m^2$). We compared our proposed method to the naïve estimate that using the subject level average of the repeated measures from activPAL and ActiGraph without measurement error correction. Participants were instructed to wear both devices at the same time

TABLE 3

*Estimated $\beta x$ in Outcome Model (BMI/Obesity) for Total Sedentary Time.*

| Outcome | BLUP$_{activPAL}$ | | BLUP$_{ActiGraph}$ | | activPAL | | ActiGraph | |
|---|---|---|---|---|---|---|---|---|
| | $\hat{\beta}_x$ (sd) | p-value | $\hat{\beta}_x$ (sd) | p-value | $\hat{\beta}_x$ (sd) | p-value | $\hat{\beta}_x$ (sd) | p-value |
| BMI | 0.80(0.09) | <0.001 | 0.77(0.10) | <0.001 | 0.55 (0.10) | <0.001 | 0.36 (0.11) | <0.001 |
| Obesity (y/n) | 0.41 (0.06) | <0.001 | 0.40 (0.06) | <0.001 | 0.28 (0.05) | <0.001 | 0.23 (0.06) | <0.001 |

for 7 days, and analysis results using BLUP from fitting LME using the activPAL measures and the ActiGraph measures are quite close.

Shown in Table 3 are the estimates of $\beta_x$ in equation (2) with associated standard error and p-values while controlling other variables that don't contain errors, such as age, gender, education level, and ethnicity. $BLUP_{activPAL}$ and $BLUP_{ActiGraph}$ indicate the proposed method using BLUP from devices activPAL and ActiGraph, while activPAL and ActiGraph indicate naïvely using measures contained errors from each device.

In Table 3, even though for both methods, the standard errors are small, and p-values < $0.001$, the estimated $\hat{\beta}_x$ in the outcome (BMI) model for total sedentary time differ substantially between the two approaches. For the proposed approach, for a 1-hour increase in the total sedentary time,average BMI is expected to increase by 0.80, while holding other covariates constant. However, naïvely using the activPAL and ActiGraph measures that contained errors gave us a falsely underestimated effect. From the naïve use of the activPAL measures, with 1-hour increase in the total sedentary time, average BMI is expected to increase by 0.55, which is about seventy percent of the estimated effect using BLUP. Meanwhile, from the naïve plug-in approach by the ActiGraph measures, with 1-hour increase in the total sedentary time, average BMI is expected to increase by 0.36, which is less than half of the estimated effect after accounting for measurement errors. Therefore without accounting for errors in the measurements, we may inappropriately underestimate the effect of sedentary time on subjects' BMI and disseminate invalid health guidance to the population.

For the binary obesity status case, we see a similar trend as in continuous BMI. For a 1hour increase in the total sedentary time, the odds of being obese will increase by around 50%
($BLUP_{activPAL}$ : exp(0.41) - 1, $BLUP_{ActiGraph}$: exp(0.40) - 1) using the proposed approach to account for the measurement errors, however, the odds of being obese will increase by around 32% (exp(0.28) - 1 ) from naïvely using activPAL measures and 26% (exp(0.23) - 1 ) from naïvely using ActiGraph measures. This falsely underestimated effect is also likely due to measurement errors, which can cause biased estimates and ultimately lead to false conclusions. Of note, both devices are also subject to random errors, and as indicated by our simulations the BLUP method may correct for both random and systematic biases in either device. Although we do not know the true $\beta_x$ for the data application, the results are consistent with our simulations where we observed shrunk estimates from the error-prone measures $W_1$ and $W_2$, while the BLUP method was able to adjust the measurement errors and give unbiased estimates.

6. Discussion. With the increasing use of wearable devices based on different technologies, there is growing interest in ascertaining how to use these rich data sources to obtain unbiased risk estimates and draw valid conclusions in health behavior studies. Although wearable sensors are likely less error-prone than self-report, however, they are subject to errors, e.g., device malfunction, or calibration issues. Notably, due to different correlation structures between measurements and errors, and other correlated variables in the model, the direction of impact on risk estimates of naïvely using measures containing errors is difficult to model or predict. Therefore, instead of estimating the impact of the measurement error, previous studies had tried to take advantage of the replicated measures to reduce the bias for estimates of the relationship between exposures subject to measurement errors and outcome of interest. For example, Rosner and Polk (1983) proposed that the average of repeated measures of subjects' blood pressure within a relatively short period of time tends to be close to their true blood pressure level. Averaging the replicates of measures may be a good idea for measurement errors that are random. However, when there are systemic correlated errors in the measurements, which can be caused by inaccurate instruments, simply using the average of the replicates may still contain a significant amount of measurement errors. Rosner et al. (1989) proposed two methods, the linear approximation method and the likelihood approximation method, to reduce bias caused by either random or systematic measurement errors for estimates of relative risk. Nonetheless, these methods require a separate validation study to be applicable. Morrel et al (2003) proposed a method using the mixed effects estimates to get a baseline measure closer to the truth than using a single measurement. However,they did not elucidate the theoretical advantages of using the BLUP, which when used correctly can be more robust to bias caused by measurement errors as demonstrated in this paper.

We developed a version of structure models by combining the Linear Mixed Effect Models and Generalized Linear Models to account for measurement errors, so that we can obtain unbiased estimations of parameters of interest. We achieved this by taking advantage of multiple replicates available from daily device wear, and proposed using the BLUP instead of naïvely using measures directly provided by the devices. We showed that using BLUP will give unbiased estimates for the conditional associations between the true exposure, i.e. sedentary time in our application, and clinical outcomes, when the true exposure is independent of other covariates that are measured without errors. Through intensive simulations for both continuous and binary outcomes, we demonstrated that the proposed method performed very well and achieved accurate parameter estimates, in scenarios with more general correlation structures and even for relatively small sample sizes. We also applied the proposed approach to an existing study and compared the results with naïve plug-in approach to demonstrate discrepancies between uncorrected estimates based on device outputs versus the error-corrected BLUP approach proposed herein.

Focusing on sedentary behavior, we compared two commonly used devices for sedentary time assessment, the activPAL and the ActiGraph. Our data analysis (Table 3) indicates less attenuation of regression coefficients for the activPAL versus ActiGraph compared to the proposed BLUP method. While we do not know the "truth" in this data analysis application, our results are in line with health behavior research: the activPAL is a validated tool for posture classification (e.g., sitting, standing, stepping), and is believed to be less biased for sedentary behavior estimation than the ActiGraph, which uses thresholds based on energy expenditure to classify sedentary behavior and thus is prone to systematic biases [28]. Of note, both devices are also subject to random errors, and as indicated by our simulations the BLUP method may correct for both random and systematic biases.

While the proposed LME-based structure models can correct the measurement errors in the exposure, it is important to note that multiple replicates are needed for the proposed method to be applicable. Since per best practices, health behavior researchers already require and collect multiple repeated days of device wear, this potential drawback, can be accommodated for sedentary behavior research, which is the focus of our application. Importantly, the LME structures can be straightforwardly implemented in different settings through standard statistical software, such as R and SAS, and generalized to other behaviors such as physical activity or sleep research.

While our approach sets a rigorous foundation, there is undoubtedly scope to expand and improve our methods. Although our data analysis study cohort was relatively complete, for future work, we are interested in expanding this work to accommodate different missingness mechanisms, such as MAR and MNAR. Meanwhile, currently we treat measures within a short period of time (7 days) as replicates of each other, we also aim to extend the current setting to longitudinal data with different cluster sizes.

# APPENDIX

FIG 5. *95 % Confidence Interval Coverage Probability of estimated $\beta_X$ using different methods for continuous outcome and binary outcome*

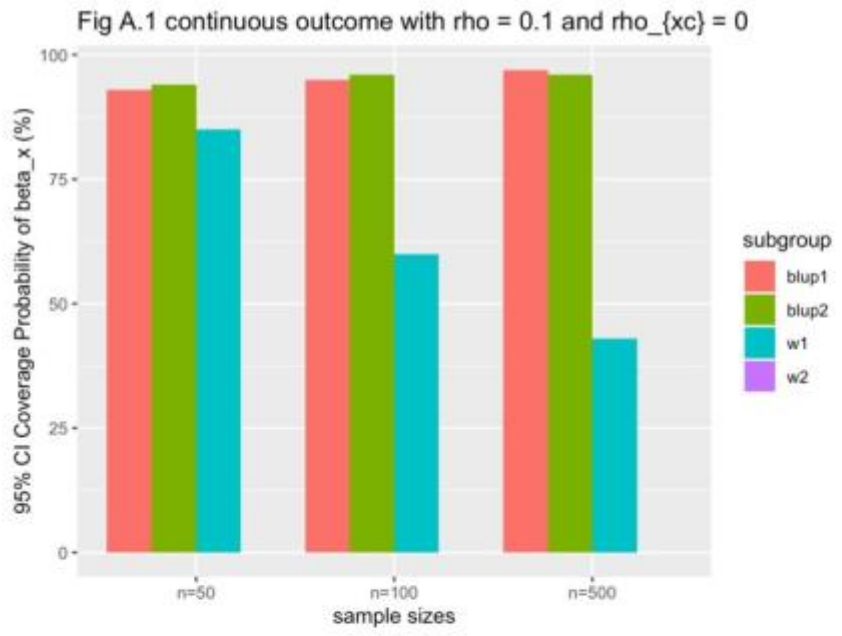

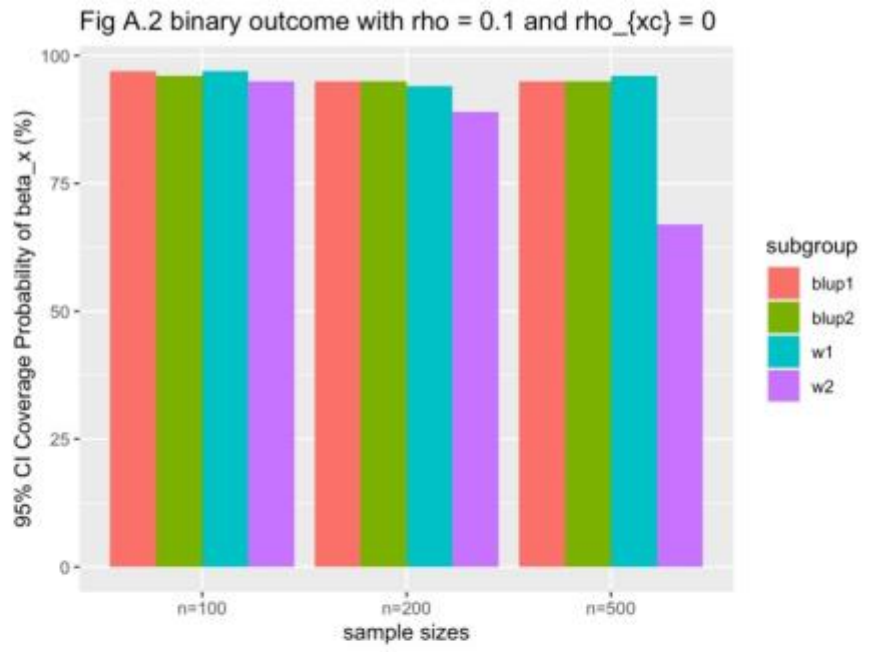

*Footnote: 95% CI coverage probability of beta_x for using w2 is 0*

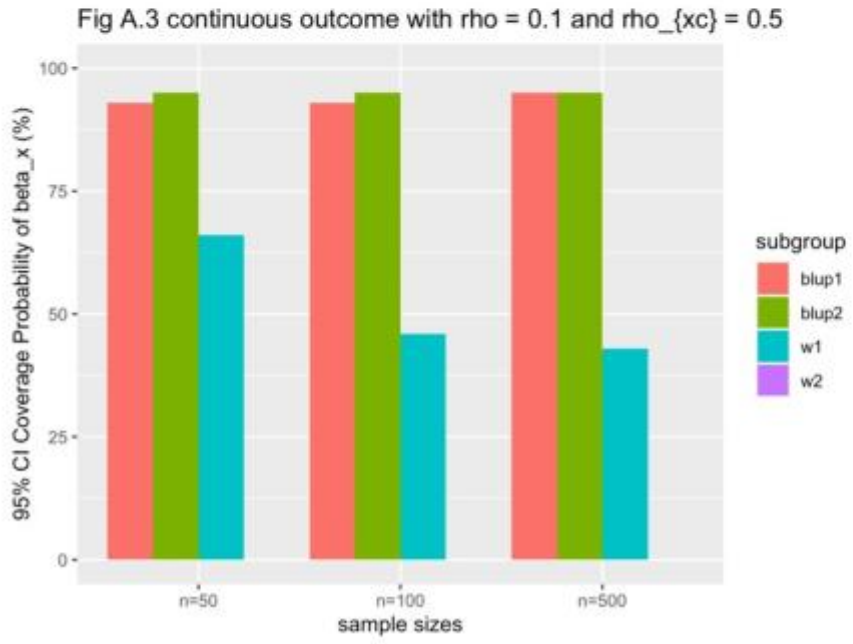

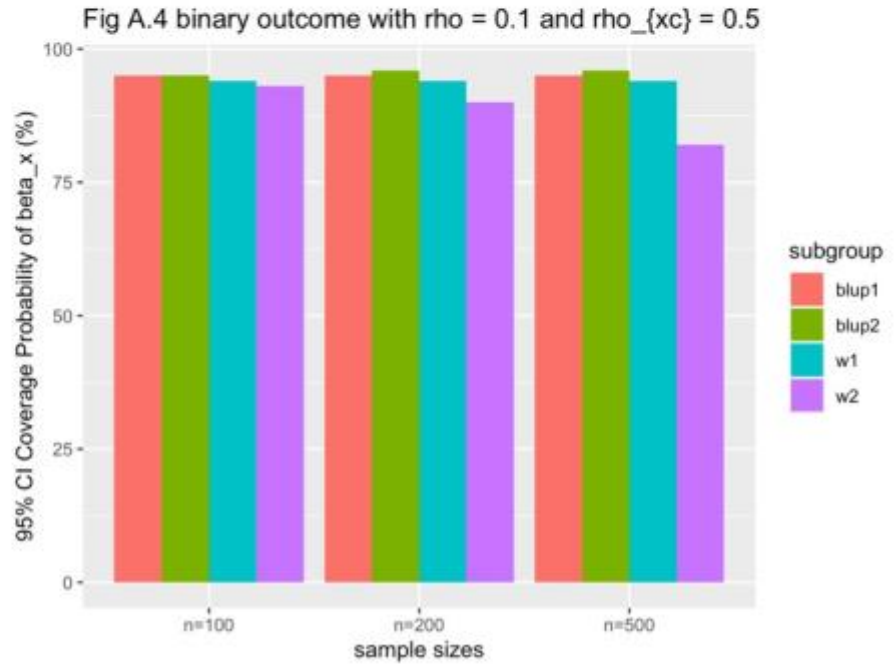

*Footnote: 95% CI coverage probability of beta_x for using w2 is 0*

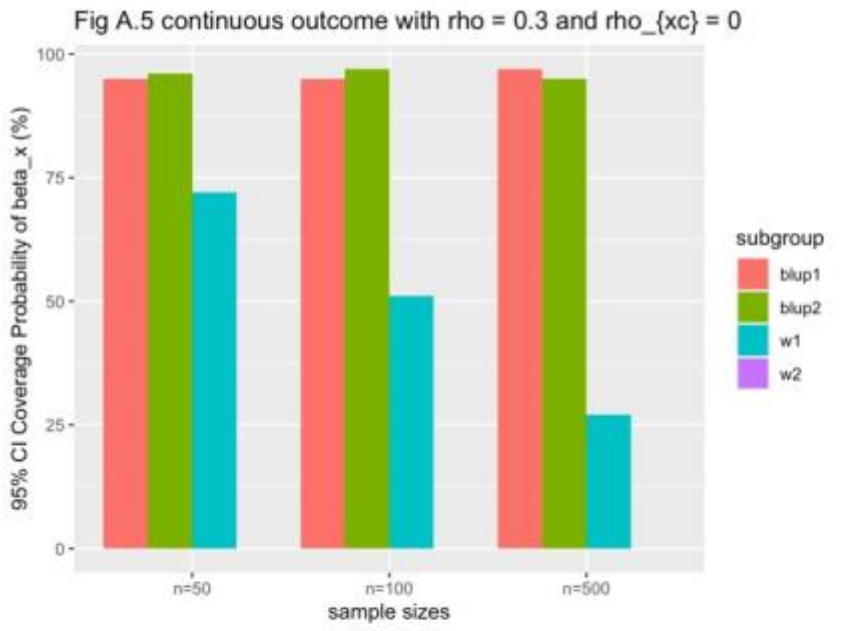

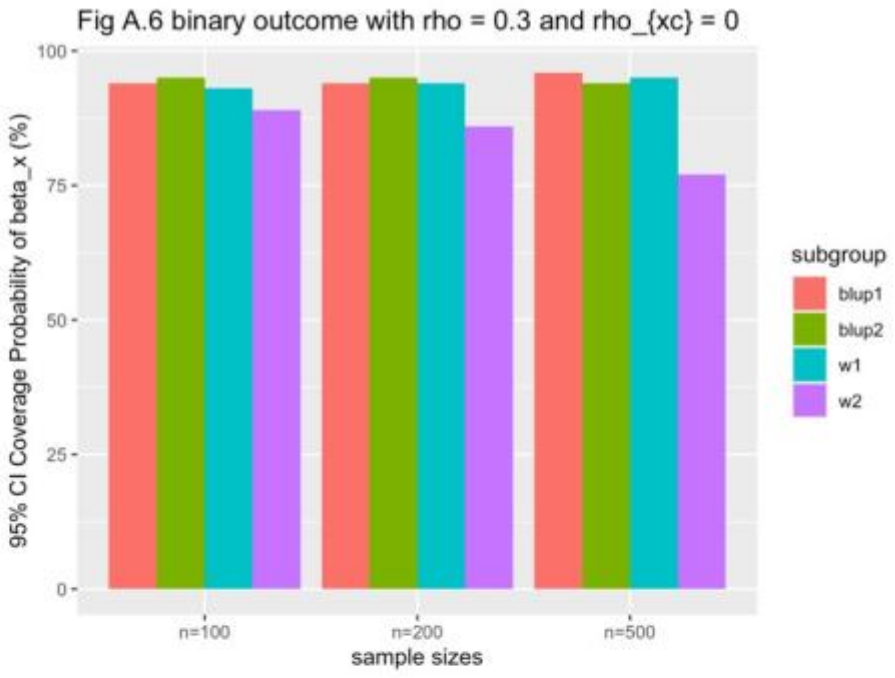

*Footnote: 95% CI coverage probability of beta_x for using w2 is 0*

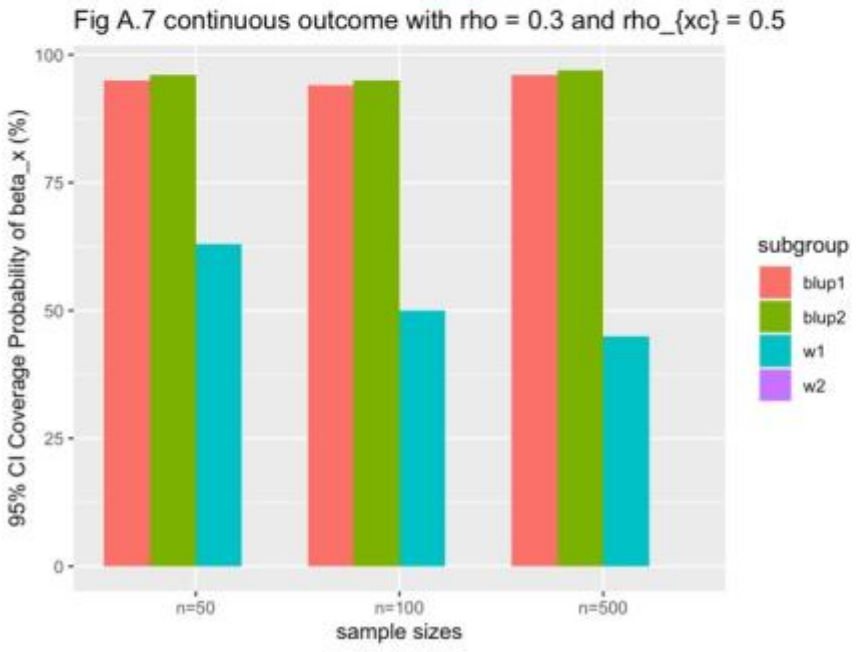

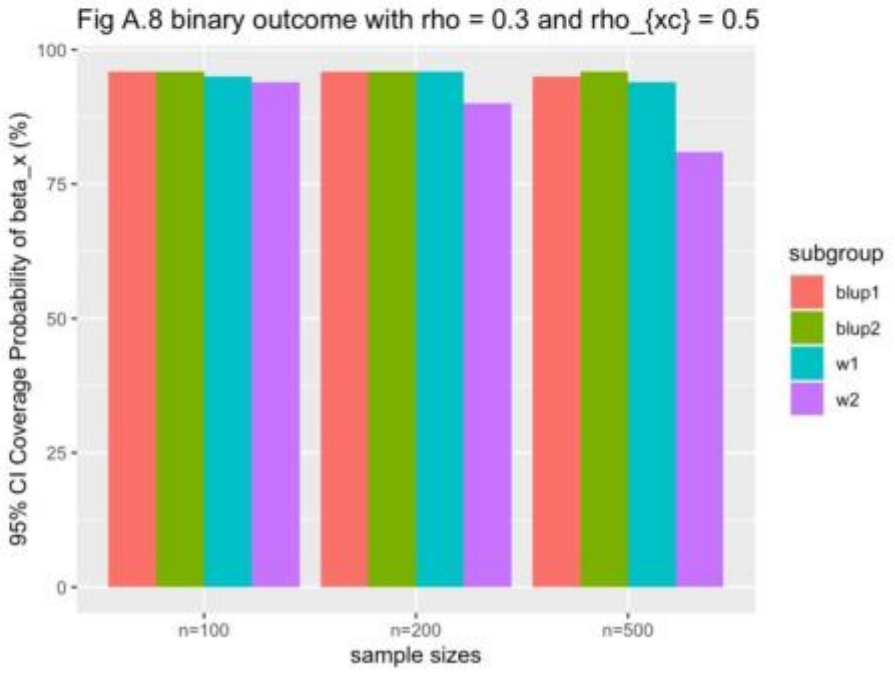

*Footnote: 95% CI coverage probability of beta_x for using w2 is 0*

TABLE
4

*Comparisons of estimates of βs under different methods (asymptotic standard error/empirical standard errors) for continuous outcome simulation study under different sample size and correlation structures.*

| | $\beta_0\,(10)$ | | | $\beta_x\,(2.95)$ | | | $\beta_c\,(3)$ | | |
|---|---|---|---|---|---|---|---|---|---|
| | n = 50 | n = 100 | n = 500 | n = 50 | n = 100 | n = 500 | n = 50 | n = 100 | n = 500 |
| **$\rho=0.1$ and $\rho_{xc}=0$** | | | | | | | | | |
| BLUP$_1$ | 10.030 (0.650/ 0.665) | 10.026 (0.329/ 0.332) | 9.986 (0.204/ 0.226) | 2.947 (0.114/ 0.115) | 2.952 (0.058/ 0.058) | 2.953 (0.036/ 0.039) | 3.002 (0.227/ 0.223) | 2.996 (0.114/ 0.105) | 3.000 (0.070/ 0.064) |
| W$_1$ | 10.539 (0.648/ 0.658) | 10.641 (0.318/ 0.305) | 10.592 (0.198/ 0.212) | 2.837 (0.113/ 0.117) | 2.828 (0.055/ 0.053) | 2.834 (0.034/ 0.035) | 3.001 (0.228/ 0.223) | 2.996 (0.114/ 0.105) | 3.000 (0.070/ 0.063) |
| BLUP$_2$ | 9.974 (0.672/ 0.729) | 9.981 (0.326/ 0.291) | 10.031 (0.204/ 0.181) | 2.953 (0.119/ 0.133) | 2.956 (0.057/ 0.052) | 2.944 (0.036/ 0.033) | 3.002 (0.228/ 0.224) | 2.998 (0.111/ 0.104) | 2.998 (0.070/ 0.070) |
| W$_2$ | 8.633 (0.533/ 0.571) | 8.719 (0.257/ 0.207) | 8.696 (0.161/ 0.146) | 1.464 (0.043/ 0.043) | 1.460 (0.021/ 0.018) | 1.460 (0.013/ 0.012) | 3.001 (0.172/ 0.173) | 2.998 (0.084/ 0.080) | 2.999 (0.053/ 0.047) |
| **$\rho=0.3$ and $\rho_{xc}=0$** | | | | | | | | | |
| BLUP$_1$ | 10.055 (0.741/ 0.823) | 9.967 (0.360/ 0.364) | 9.985 (0.228/ 0.241) | 2.949 (0.130/ 0.139) | 2.952 (0.063/ 0.068) | 2.955 (0.040/ 0.042) | 2.997 (0.254/ 0.237) | 3.002 (0.123/ 0.124) | 2.998 (0.077/ 0.078) |
| W$_1$ | 10.905 (0.707/ 0.746) | 10.816 (0.344/ 0.324) | 10.809 (0.218/ 0.220) | 2.778 (0.122/ 0.123) | 2.782 (0.059/ 0.060) | 2.784 (0.038/ 0.037) | 2.996 (0.254/ 0.236) | 3.003 (0.123/ 0.124) | 2.998 (0.077/ 0.078) |
| BLUP$_2$ | 9.980 (0.752/ 0.871) | 9.974 (0.366/ 0.418) | 9.994 (0.228/ 0.260) | 2.951 (0.132/ 0.148) | 2.955 (0.064/ 0.072) | 2.955 (0.040/ 0.045) | 2.999 (0.250/ 0.263) | 2.997 (0.124/ 0.131) | 2.998 (0.078/ 0.083) |
| W$_2$ | 8.828 (0.554/ 0.553) | 8.747 (0.269/ 0.327) | 8.756 (0.169/ 0.182) | 1.453 (0.045/ 0.044) | 1.452 (0.022/ 0.027) | 1.453 (0.014/ 0.015) | 2.992 (0.182/ 0.166) | 3.001 (0.088/ 0.093) | 2.995 (0.055/ 0.058) |
| **$\rho=0.1$ and $\rho_{xc}=0.5$** | | | | | | | | | |
| BLUP$_1$ | 9.932 (0.619/ 0.686) | 10.035 (0.305/ 0.302) | 10.029 (0.193/ 0.210) | 2.920 (0.132/ 0.150) | 2.929 (0.065/ 0.066) | 2.942 (0.041/ 0.043) | 3.196 (0.259/ 0.296) | 3.169 (0.127/ 0.139) | 3.162 (0.080/ 0.079) |
| W$_1$ | 10.564 (0.594/ 0.611) | 10.648 (0.293/ 0.279) | 10.637 (0.186/ 0.190) | 2.793 (0.126/ 0.132) | 2.787 (0.062/ 0.061) | 2.790 (0.039/ 0.039) | 3.196 (0.259/ 0.296) | 3.169 (0.127/ 0.139) | 3.162 (0.080/ 0.079) |
| BLUP$_2$ | 9.951 (0.620/ 0.695) | 10.013 (0.307/ 0.313) | 10.027 (0.193/ 0.190) | 2.935 (0.134/ 0.122) | 2.937 (0.065/ 0.065) | 2.944 (0.041/ 0.041) | 3.138 (0.259/ 0.263) | 3.155 (0.126/ 0.127) | 3.158 (0.080/ 0.078) |
| W$_2$ | 8.679 (0.496/ 0.459) | 8.709 (0.245/ 0.235) | 8.713 (0.154/ 0.151) | 1.450 (0.048/ 0.045) | 1.454 (0.024/ 0.023) | 1.454 (0.015/ 0.015) | 3.193 (0.196/ 0.215) | 3.244 (0.096/ 0.099) | 3.242 (0.060/ 0.060) |
| **$\rho=0.3$ and $\rho_{xc}=0.5$** | | | | | | | | | |
| BLUP$_1$ | 9.976 (0.705/ 0.730) | 10.039 (0.337/ 0.351) | 10.041 (0.214/ 0.238) | 2.922 (0.149/ 0.148) | 2.933 (0.072/ 0.075) | 2.947 (0.045/ 0.051) | 3.238 (0.288/ 0.281) | 3.221 (0.140/ 0.144) | 3.217 (0.088/ 0.087) |
| W$_1$ | 10.870 (0.665/ 0.642) | 10.864 (0.319/ 0.315) | 10.875 (0.202/ 0.215) | 2.725 (0.140/ 0.130) | 2.732 (0.067/ 0.068) | 2.731 (0.043/ 0.046) | 3.238 (0.288/ 0.281) | 3.221 (0.139/ 0.144) | 3.217 (0.088/ 0.087) |
| BLUP$_2$ | 9.989 (0.695/ 0.723) | 10.045 (0.338/ 0.350) | 10.053 (0.214/ 0.221) | 2.931 (0.148/ 0.151) | 2.939 (0.072/ 0.075) | 2.951 (0.045/ 0.047) | 3.213 (0.287/ 0.295) | 3.228 (0.139/ 0.134) | 3.219 (0.088/ 0.086) |
| W$_2$ | 8.797 (0.529/ 0.521) | 8.789 (0.257/ 0.256) | 8.787 (0.162/ 0.164) | 1.445 (0.051/ 0.052) | 1.446 (0.025/ 0.025) | 1.446 (0.016/ 0.016) | 3.258 (0.207/ 0.208) | 3.256 (0.101/ 0.104) | 3.256 (0.063/ 0.066) |

5

*Comparisons of estimates of βs under different methods (asymptotic standard error/empirical standard errors) for binary outcome simulation study under different sample size and correlation structures.*

TABLE

| | $\beta_0\,(0.1)$ | | | $\beta_x\,(0.1)$ | | | $\beta_c\,(0.1)$ | | |
|---|---|---|---|---|---|---|---|---|---|
| | n = 100 | n = 200 | n = 500 | n = 100 | n = 200 | n = 500 | n = 100 | n = 200 | n = 500 |
| **$\rho = 0.1$ and $\rho_{xc} = 0$** | | | | | | | | | |
| $BLUP_1$ | 0.108 (0.350/0.349) | 0.097 (0.260/0.274) | 0.097 (0.120/0.117) | 0.099 (0.190/0.183) | 0.106 (0.160/0.167) | 0.105 (0.074/0.071) | 0.105 (0.212/0.225) | 0.109 (0.147/0.150) | 0.102 (0.068/0.063) |
| $W_1$ | 0.083 (0.379/0.376) | 0.113 (0.246/0.258) | 0.112 (0.114/0.111) | 0.124 (0.239/0.231) | 0.090 (0.136/0.141) | 0.089 (0.063/0.061) | 0.105 (0.212/0.225) | 0.109 (0.147/0.150) | 0.102 (0.068/0.063) |
| $BLUP_2$ | 0.109 (0.359/0.378) | 0.098 (0.261/0.278) | 0.096 (0.121/0.120) | 0.103 (0.195/0.204) | 0.104 (0.159/0.165) | 0.101 (0.074/0.075) | 0.105 (0.213/0.226) | 0.108 (0.149/0.151) | 0.102 (0.068/0.063) |
| $W_2$ | 0.030 (0.420/0.426) | 0.053 (0.297/0.315) | 0.058 (0.138/0.139) | 0.059 (0.101/0.103) | 0.050 (0.072/0.071) | 0.048 (0.033/0.034) | 0.106 (0.212/0.225) | 0.109 (0.147/0.149) | 0.102 (0.068/0.063) |
| **$\rho = 0.3$ and $\rho_{xc} = 0$** | | | | | | | | | |
| $BLUP_1$ | 0.103 (0.378/0.402) | 0.111 (0.262/0.278) | 0.099 (0.164/0.162) | 0.098 (0.238/0.263) | 0.103 (0.164/0.180) | 0.096 (0.102/0.102) | 0.107 (0.213/0.222) | 0.097 (0.147/0.149) | 0.098 (0.092/0.092) |
| $W_1$ | 0.120 (0.349/0.368) | 0.131 (0.243/0.256) | 0.118 (0.152/0.151) | 0.071 (0.190/0.208) | 0.083 (0.132/0.145) | 0.077 (0.082/0.081) | 0.107 (0.213/0.222) | 0.097 (0.147/0.149) | 0.098 (0.092/0.092) |
| $BLUP_2$ | 0.104 (0.373/0.398) | 0.094 (0.262/0.265) | 0.102 (0.164/0.165) | 0.093 (0.240/0.281) | 0.103 (0.165/0.169) | 0.103 (0.102/0.106) | 0.107 (0.213/0.222) | 0.097 (0.147/0.151) | 0.098 (0.092/0.092) |
| $W_2$ | 0.060 (0.425/0.452) | 0.071 (0.295/0.308) | 0.052 (0.185/0.186) | 0.043 (0.103/0.133) | 0.048 (0.071/0.074) | 0.047 (0.044/0.045) | 0.107 (0.213/0.223) | 0.097 (0.147/0.149) | 0.098 (0.092/0.092) |
| **$\rho = 0.1$ and $\rho_{xc} = 0.5$** | | | | | | | | | |
| $BLUP_1$ | 0.103 (0.316/0.329) | 0.097 (0.229/0.230) | 0.097 (0.143/0.146) | 0.098 (0.262/0.274) | 0.095 (0.180/0.181) | 0.097 (0.112/0.115) | 0.115 (0.239/0.240) | 0.115 (0.165/0.173) | 0.116 (0.104/0.110) |
| $W_1$ | 0.088 (0.333/0.347) | 0.112 (0.218/0.219) | 0.088 (0.137/0.139) | 0.083 (0.221/0.231) | 0.081 (0.153/0.153) | 0.082 (0.096/0.098) | 0.115 (0.239/0.240) | 0.115 (0.165/0.173) | 0.116 (0.104/0.110) |
| $BLUP_2$ | 0.102 (0.323/0.331) | 0.099 (0.230/0.231) | 0.095 (0.144/0.149) | 0.101 (0.261/0.271) | 0.096 (0.180/0.181) | 0.101 (0.113/0.117) | 0.115 (0.239/0.240) | 0.115 (0.165/0.173) | 0.116 (0.104/0.110) |
| $W_2$ | 0.053 (0.380/0.390) | 0.060 (0.263/0.266) | 0.043 (0.164/0.164) | 0.046 (0.119/0.118) | 0.046 (0.083/0.083) | 0.048 (0.052/0.052) | 0.115 (0.243/0.244) | 0.115 (0.168/0.175) | 0.118 (0.105/0.113) |
| **$\rho = 0.3$ and $\rho_{xc} = 0.5$** | | | | | | | | | |
| $BLUP_1$ | 0.095 (0.318/0.351) | 0.096 (0.232/0.236) | 0.099 (0.145/0.153) | 0.094 (0.266/0.284) | 0.105 (0.184/0.185) | 0.097 (0.115/0.118) | 0.123 (0.237/0.254) | 0.113 (0.164/0.164) | 0.114 (0.103/0.106) |
| $W_1$ | 0.079 (0.313/0.312) | 0.113 (0.216/0.221) | 0.109 (0.136/0.137) | 0.085 (0.225/0.279) | 0.084 (0.157/0.157) | 0.078 (0.092/0.094) | 0.123 (0.237/0.254) | 0.113 (0.164/0.164) | 0.114 (0.103/0.106) |
| $BLUP_2$ | 0.097 (0.320/0.359) | 0.098 (0.233/0.240) | 0.096 (0.145/0.145) | 0.102 (0.269/0.291) | 0.103 (0.185/0.185) | 0.095 (0.115/0.115) | 0.123 (0.237/0.254) | 0.113 (0.164/0.164) | 0.115 (0.103/0.106) |
| $W_2$ | 0.021 (0.372/0.377) | 0.040 (0.261/0.266) | 0.051 (0.163/0.164) | 0.048 (0.118/0.118) | 0.050 (0.082/0.083) | 0.048 (0.051/0.051) | 0.122 (0.243/0.259) | 0.116 (0.167/0.170) | 0.115 (0.105/0.109) |

6

*Comparisons of relative bias and coverage probability for estimates of $\beta$s under different methods for continuous outcome simulation study under different sample size and correlation structures.*

$\rho = 0.1$ and $\rho_{xc} = 0$

TABLE

| | $\beta_0$ | | | $\beta_x$ | | | $\beta_c$ | | |
|---|---|---|---|---|---|---|---|---|---|
| | n = 50 | n = 100 | n = 500 | n = 50 | n = 100 | n = 500 | n = 50 | n = 100 | n = 500 |
| relative bias BLUP$_1$ | 0.30% | 0.26% | 0.14% | 0.10% | 0.07% | 0.10% | 0.06% | 0.13% | 0 |
| coverage BLUP$_1$ | 89% | 95% | 97% | 93% | 95% | 97% | 96% | 97% | 99% |
| relative bias W$_1$ | 5.39% | 6.41% | 5.92% | 3.83% | 4.14% | 3.93% | 0.03% | 0.13% | 0 |
| coverage W$_1$ | 85% | 70% | 11% | 85% | 60% | 43% | 95% | 96% | 99% |
| relative bias BLUP$_2$ | 0.26% | 0.19% | 0.31% | 0.10% | 0.20% | 0.20% | 0.06% | 0.06% | 0.06% |
| coverage BLUP$_2$ | 95% | 94% | 97% | 94% | 96% | 96% | 93% | 97% | 99% |
| relative bias W$_2$ | 13.67% | 12.81% | 13.04% | 50.37% | 50.51% | 50.51% | 0.03% | 0.06% | 0.03% |
| coverage W$_2$ | 25% | 8% | 0 | 0 | 0 | 0 | 97% | 95% | 99% |
| $\rho = 0.3$ and $\rho_{xc} = 0$ | | | | | | | | | |
| relative bias BLUP$_1$ | 0.55% | 0.33% | 0.15% | 0.03% | 0.07% | 0.17% | 0.10% | 0.06% | 0.06% |
| coverage BLUP$_1$ | 90% | 97% | 94% | 95% | 95% | 97% | 96% | 97% | 95% |
| relative bias W$_1$ | 9.05% | 8.16% | 8.09% | 5.83% | 5.69% | 5.63% | 0.13% | 0.10% | 0.06% |
| coverage W$_1$ | 80% | 63% | 18% | 72% | 51% | 27% | 95% | 96% | 94% |
| relative bias BLUP$_2$ | 0.20% | 0.26% | 0.06% | 0.03% | 0.17% | 0.17% | 0.03% | 0.10% | 0.06% |
| coverage BLUP$_2$ | 92% | 94% | 95% | 96% | 97% | 95% | 96% | 97% | 96% |
| relative bias W$_2$ | 11.72% | 12.53% | 12.44% | 50.75% | 50.78% | 50.75% | 0.27% | 0.03% | 0.17% |
| coverage W$_2$ | 39% | 9% | 0 | 0 | 0 | 0 | 93% | 96% | 98% |
| $\rho = 0.1$ and $\rho_{xc} = 0.5$ | | | | | | | | | |
| relative bias BLUP$_1$ | 0.68% | 0.35% | 0.29% | 1.02% | 0.71% | 0.27% | 6.53% | 5.63% | 5.40% |
| coverage BLUP$_1$ | 95% | 93% | 94% | 93% | 93% | 95% | 86% | 85% | 79% |
| relative bias W$_1$ | 5.64% | 6.48% | 6.37% | 5.32% | 5.53% | 5.42% | 6.53% | 5.63% | 5.40% |
| coverage W$_1$ | 74% | 58% | 29% | 66% | 46% | 43% | 83% | 81% | 77% |
| relative bias BLUP$_2$ | 0.49% | 0.13% | 0.27% | 0.51% | 0.44% | 0.20% | 4.60% | 5.17% | 5.27% |
| coverage BLUP$_2$ | 96% | 94% | 95% | 95% | 95% | 95% | 85% | 84% | 81% |
| relative bias W$_2$ | 13.21% | 12.91% | 12.87% | 50.85% | 50.71% | 50.71% | 6.43% | 8.13% | 8.07% |
| coverage W$_2$ | 20% | 19% | 0 | 0 | 0 | 0 | 88% | 89% | 80% |
| $\rho = 0.3$ and $\rho_{xc} = 0.5$ | | | | | | | | | |
| relative bias BLUP$_1$ | 0.24% | 0.39% | 0.41% | 0.95% | 0.58% | 0.44% | 7.93% | 7.37% | 7.23% |
| coverage BLUP$_1$ | 95% | 94% | 96% | 95% | 94% | 96% | 87% | 83% | 76% |
| relative bias W$_1$ | 8.70% | 8.64% | 8.75% | 7.63% | 7.39% | 7.42% | 7.93% | 7.37% | 7.23% |
| coverage W$_1$ | 69% | 61% | 31% | 63% | 50% | 45% | 86% | 80% | 76% |
| relative bias BLUP$_2$ | 0.11% | 0.45% | 0.53% | 0.64% | 0.37% | 0.31% | 7.10% | 7.60% | 7.30% |
| coverage BLUP$_2$ | 94% | 93% | 95% | 96% | 95% | 97% | 88% | 82% | 77% |
| relative bias W$_2$ | 12.03% | 12.11% | 12.13% | 51.02% | 50.98% | 50.98% | 8.60% | 8.53% | 8.53% |
| coverage W$_2$ | 27% | 12% | 0 | 0 | 0 | 0 | 90% | 86% | 78% |

7

*Comparisons of relative bias and coverage probability for estimates of βs under different methods for binary outcome simulation study under different sample size and correlation structures.*

$\rho = 0.1$ and $\rho_{xc} = 0$

| | $\beta_0$ | | | $\beta_x$ | | | $\beta_c$ | | |
|---|---|---|---|---|---|---|---|---|---|
| | n = 100 | n = 200 | n = 500 | n = 100 | n = 200 | n = 500 | n = 100 | n = 200 | n = 500 |
| relative bias BLUP$_1$ | 7.81% | 3.00% | 3.23% | 0.80% | 5.69% | 5.12% | 5.36% | 9.33% | 1.97% |

TABLE

| | | | | | | | | | |
|---|---|---|---|---|---|---|---|---|---|
| coverage $BLUP_1$ | 96% | 94% | 96% | 97% | 95% | 95% | 95% | 96% | 96% |
| relative bias $W_1$ | 17.36% | 13.21% | 12.40% | 24.38% | 10.15% | 10.50% | 5.36% | 9.32% | 1.97% |
| coverage $W_1$ | 96% | 94% | 95% | 97% | 94% | 96% | 95% | 95% | 96% |
| relative bias $BLUP_2$ | 9.00% | 2.00% | 4.00% | 3.00% | 4.00% | 1.00% | 5.36% | 8.00% | 1.97% |
| coverage $BLUP_2$ | 95% | 95% | 96% | 96% | 95% | 95% | 95% | 95% | 96% |
| relative bias $W_2$ | 70.41% | 47.41% | 41.88% | 41.18% | 49.85% | 52.08% | 6.07% | 9.04% | 2.00% |
| coverage $W_2$ | 96% | 94% | 94% | 95% | 89% | 67% | 94% | 95% | 96% |
| $\rho = 0.3$ and $\rho_{xc} = 0$ | | | | | | | | | |
| relative bias $BLUP_1$ | 3.11% | 11.20% | 0.87% | 2.00% | 2.53% | 4.54% | 7.23% | 2.79% | 2.30% |
| coverage $BLUP_1$ | 95% | 94% | 95% | 94% | 94% | 96% | 96% | 97% | 96% |
| relative bias $W_1$ | 20.10% | 31.14% | 17.76% | 28.97% | 17.38% | 23.17% | 7.26% | 2.77% | 2.30% |
| coverage $W_1$ | 95% | 94% | 95% | 93% | 94% | 95% | 96% | 97% | 95% |
| relative bias $BLUP_2$ | 4.00% | 5.93% | 1.80% | 7.00% | 2.54% | 3.81% | 7.23% | 2.78% | 2.30% |
| coverage $BLUP_2$ | 95% | 94% | 95% | 95% | 95% | 94% | 96% | 97% | 96% |
| relative bias $W_2$ | 39.59% | 29.18% | 47.61% | 56.62% | 52.39% | 52.59% | 6.80% | 2.84% | 2.40% |
| coverage $W_2$ | 95% | 94% | 94% | 89% | 86% | 77% | 95% | 97% | 96% |
| $\rho = 0.1$ and $\rho_{xc} = 0.5$ | | | | | | | | | |
| relative bias $BLUP_1$ | 3.19% | 2.76% | 2.58% | 2.06% | 4.63% | 3.38% | 15.11% | 14.71% | 16.25% |
| coverage $BLUP_1$ | 94% | 96% | 95% | 95% | 95% | 95% | 96% | 96% | 94% |
| relative bias $W_1$ | 12.48% | 11.57% | 12.37% | 17.39% | 18.89% | 18.42% | 15.11% | 14.70% | 16.26% |
| coverage $W_1$ | 94% | 95% | 95% | 94% | 94% | 94% | 96% | 95% | 94% |
| relative bias $BLUP_2$ | 2.11% | 0.25% | 4.86% | 1.16% | 4.26% | 1.42% | 15.11% | 14.71% | 16.25% |
| coverage $BLUP_2$ | 95% | 96% | 95% | 95% | 96% | 96% | 96% | 96% | 94% |
| relative bias $W_2$ | 47.11% | 39.66% | 57.32% | 53.97% | 54.10% | 52.02% | 9.50% | 14.69% | 18.13% |
| coverage $W_2$ | 95% | 95% | 94% | 93% | 90% | 82% | 96% | 96% | 94% |
| $\rho = 0.3$ and $\rho_{xc} = 0.5$ | | | | | | | | | |
| relative bias $BLUP_1$ | 5.06% | 4.12% | 1.07% | 6.01% | 4.95% | 2.68% | 22.76% | 12.93% | 14.07% |
| coverage $BLUP_1$ | 96% | 95% | 95% | 96% | 96% | 95% | 96% | 96% | 96% |
| relative bias $W_1$ | 20.62% | 13.32% | 8.57% | 15.04% | 15.55% | 21.78% | 22.81% | 12.95% | 14.06% |
| coverage $W_1$ | 96% | 95% | 96% | 95% | 96% | 94% | 96% | 96% | 95% |
| relative bias $BLUP_2$ | 3.19% | 2.61% | 3.87% | 2.35% | 3.93% | 5.26% | 22.76% | 12.93% | 14.92% |
| coverage $BLUP_2$ | 96% | 96% | 96% | 96% | 96% | 96% | 96% | 96% | 96% |
| relative bias $W_2$ | 78.74% | 60.32% | 48.57% | 52.00% | 50.26% | 51.84% | 22.43% | 16.38% | 15.08% |
| coverage $W_2$ | 96% | 95% | 94% | 94% | 90% | 81% | 95% | 95% | 94% |

Acknowledgments. We would like to thank the anonymous referees, an Associate Editor and the Editor for their constructive comments that improved the quality of this paper. We would also like to thank the participants of the Adult Changes in Thought (ACT) study for the data they have provided and the many ACT investigators and staff who steward that data.

Funding. SJH, ALC, CD, and LN were partially supported by funding from the National Institute of Aging (PO1AG052352); ALC and LN were partially supported by funding from the National Institute of Diabetes and Digestive and Kidney Disease (R01DK114945); CD and LN were partially supported by funding from the National Heart, Lung and Blood Institute (R01HL130483). This work's data collection was additionally supported by prior funding from the National Institute on Aging (U01AG006781). All statements in this report, including its findings and conclusions, are solely those of the authors and do not necessarily represent the views of the National Institute on Aging or the National Institutes of Health.

References.